\documentclass[11pt]{article}
\usepackage{amssymb}
\usepackage{amsmath}
\usepackage{amscd}
\usepackage{bbm}
\usepackage{latexsym}
\usepackage{ifthen}
\usepackage[xdvi]{graphics}

\catcode`@=11
\renewcommand{\section}{\@startsection{section}{1}{0pt}{\medskipamount}
{\medskipamount}{\large\bf}}
\numberwithin{equation}{section}
\catcode`@=12

\def\a{\alpha}

\def\de{\delta}

\def\ve{\varepsilon}

\def\h{\eta}

\def\s{\sigma}

\def\vp{\varphi}

\def\ps{\psi}
\def\om{\omega}
\def\t{\tau}
\def\P{\Phi}

\def\S{\Sigma}

\newcommand{\Om}{\Omega} 
\newcommand{\C}{\mathbbm{C}}
\newcommand{\R}{\mathbbm{R}}
\newcommand{\Z}{\mathbbm{Z}}
\newcommand{\Na}{\mathbbm{N}}

\newcommand{\Hcal}{{\cal H}}

\newcommand{\Ical}{{\cal I}}

\def\>{\rangle}
\def\<{\langle}
\def\N2{N=2}
\def\pa{\partial}

\def\tr{{\rm tr}}
\def\sfrac#1#2{{\textstyle\frac{#1}{#2}}}

\newcommand{\ab}{{\bar{a}}}
\newcommand{\bb}{{\bar{b}}}

\newcommand{\nb}{{\bar{0}}}
\newcommand{\ob}{{\bar{1}}}

\newcommand{\Zb}{\bar{Z}}

\newcommand{\zb}{\bar{z}}

\newcommand{\pab}{\bar{\pa}}

\newcommand{\Gt}{\widetilde{G}}

\newcommand{\Phh}{{\widehat{\Phi}}}

\newcommand{\ic}{\text{i}}

\newcommand{\Gp}{\ifthenelse{\boolean{mmode}}{{G^+}}{\mbox{$G^+\:$}}}
\newcommand{\Gtp}{\ifthenelse{\boolean{mmode}}{\mbox{$\Gt^+$}}{\mbox{$\Gt^+\:$}}}
\newcommand{\Gm}{\ifthenelse{\boolean{mmode}}{{G^-}}{\mbox{$G^-\:$}}}
\newcommand{\Gtm}{\ifthenelse{\boolean{mmode}}{\mbox{$\Gt^-$}}{\mbox{$\Gt^-\:$}}}

\newcommand{\uv}{\text{$|\;\!\!\!\uparrow\>$}}
\newcommand{\dv}{\text{$|\;\!\!\!\downarrow\>$}}
\newcommand{\uvb}{\text{$\<\uparrow\:\!\!\!|$}}
\newcommand{\dvb}{\text{$\<\downarrow\:\!\!\!|$}}

\newcommand{\udv}{|\;\!\!\!\uparrow\;\!\!\downarrow\>}
\newcommand{\duv}{|\;\!\!\!\downarrow\;\!\!\uparrow\>}

\newcommand{\udvb}{\<\uparrow\downarrow\;\!\!\!|}
\newcommand{\duvb}{\<\downarrow\uparrow\;\!\!\!|}

\newcommand{\uudvb}{\<\uparrow\uparrow\downarrow\;\!\!\!|}
\newcommand{\uduvb}{\<\uparrow\downarrow\uparrow\;\!\!\!|}
\newcommand{\duuvb}{\<\downarrow\uparrow\uparrow\;\!\!\!|}
\newcommand{\bpz}{\text{bpz}}
\newcommand{\stl}[1]{{}_{#1}^{\phantom{\dagger}}}
\newcommand{\sap}{\text{\raisebox{-.5mm}{$\sqrt{{\a'}}$}}}

\begin{document}
\begin{titlepage}
\setcounter{page}{0}
\begin{flushright}
{\tt hep-th/0306254}\\
ITP--UH--03/03\\
\end{flushright}

\vskip 2.0cm

\begin{center}

{\Large\bf String Field Theory Vertices for Fermions of Integral Weight}

\vspace{14mm}

{\large Alexander Kling \ and \ Sebastian Uhlmann}
\\[5mm]
{\em Institut f\"ur Theoretische Physik  \\
Universit\"at Hannover \\
Appelstra\ss{}e 2, 30167 Hannover, Germany }\\[2mm]
{E-mail: {\tt kling, uhlmann@itp.uni-hannover.de}}

\end{center}

\vspace{2cm}

\noindent
{\sc Abstract:} \\[.7cm]
We construct Witten-type string field theory vertices for a fermionic 
first order system with conformal weights $(0,1)$ in the operator formulation 
using delta-function overlap conditions as well as the Neumann function 
method. The identity, the reflector and the interaction vertex are treated in 
detail paying attention to the zero mode conditions and the $U(1)$ charge 
anomaly. The Neumann coefficients for the interaction vertex are shown to be 
intimately connected with the coefficients for bosons allowing a simple proof 
that the reparametrization anomaly of the fermionic first order system 
cancels the contribution of two real bosons. This agrees with their 
contribution $c=-2$ to the central charge. The overlap equations for the 
interaction vertex are shown to hold. Our results have applications in 
N=2 string field theory, Berkovits' hybrid formalism for superstring field 
theory, the $\eta\xi$-system and the twisted $bc$-system used in bosonic 
vacuum string field theory.

\vfill

\textwidth 6.5truein

\end{titlepage}


\section{Introduction}
\noindent
The study of nonperturbative physics in string theory has enjoyed a lively 
interest in recent years. String field theory, especially in the formulation 
of~\cite{Witten:1985cc}, has proven to be a useful tool for developing 
qualitative as well as quantitative results. In particular, Sen's conjectures 
on tachyon condensation~\cite{Sen:1999mh,Sen:1999mg,Sen:1999xm} have set the 
stage for putting string field theory to work (see, for instance, the 
reviews~\cite{Ohmori:2001am,DeSmet:2001af,Arefeva:2001ps,Taylor:2002uv} and 
references therein). In this framework it should be possible to
describe the result of the condensation process, the closed string vacuum,
as a solution to its equation of motion. However, no plausible candidate
for a solution in open string field theory has been found yet.

In order to gain some insight into the structure of solutions describing 
D-branes, Rastelli, Sen and Zwiebach proposed to expand all string fields 
around a (unknown) closed string vacuum solution. This vacuum 
string field theory~\cite{Rastelli:2000hv,Rastelli:2001jb} is
described as a certain singular limit, in which the kinetic operator 
consists only of ghosts. 
In this limit, the equation of motion of bosonic string field theory 
factorizes into a matter and a ghost part, which can be solved independently; 
the matter part of the string field has to be a projector of the star algebra. 
This separation, however, does not pertain to open string field theory, where 
the kinetic operator mixes matter and ghost sectors. Much work has been done 
in the context of bosonic string theory to classify these solutions in 
terms of projectors of the star algebra~\cite{Rastelli:2001rj,Gross:2001rk,%
Gross:2001yk,Schnabl:2002ff,Gaiotto:2002kf,Fuchs:2002zz}, 
most prominently the sliver. They have been identified with (multiple) 
D-branes of various dimensions, and it was shown that they reproduce the 
correct ratio of tensions. Their similarity to noncommutative solitons has 
been investigated~\cite{Chen:2002jd,Bonora:2002rn}. In such considerations 
a major technical simplification is achieved by taking advantage of the map 
from Witten's star product to a continuous\footnote{Originally, the discrete 
version of this map has been proposed in~\cite{Bars:2001ag} and further 
developed in~\cite{Bars:2002nu}.} Moyal product~\cite{Douglas:2002jm}. 

Much less is known for superstrings. Different generalizations of 
bosonic vacuum string field theory to non-BPS and $D$-$\bar D$ brane systems 
have been proposed~\cite{Kluson:2001sb,Marino:2001ny,Arefeva:2002mb} using 
Berkovits' formulation for a nonpolynomial superstring field 
theory~\cite{Berkovits:1995ab} as well as cubic open superstring field 
theory~\cite{Witten:1986qs,Arefeva:1989cp}. Although superstring field 
theories applied to the problem of tachyon condensation perform 
well~\cite{Berkovits:2000hf,DeSmet:2000dp,Aref'eva:2000mb}, it is considerably 
harder to make sense of the conjectured versions around the tachyon vacuum. 
Despite considerable 
efforts~\cite{Kluson:2002kk,Arefeva:2002sg,Ohmori:2002ah,Ohmori:2002kj} 
up to now no satisfactory solutions to vacuum 
superstring field theory have been given. Even worse, recent results using 
the numerical technique of level truncation seem to indicate that 
the pure ghost ansatz for the kinetic operator fails to describe the theory 
around the tachyon vacuum~\cite{Ohmori:2003vq}. Therefore it is necessary to 
gain more insight into the structure of solutions to string field theory with 
more general kinetic operators mixing different sectors of the theory. This 
is, of course, a technically demanding task.

Due to these difficulties it seems worthwhile to take an apparent sidestep and 
to consider alternative approaches to tackle the problems stated 
above. In order to study the properties of string field theory solutions with
these mixing properties, one may study a different string field theory
containing world-sheet fermions instead of reparametrization ghosts.
Namely, string field theory for N=2 strings~\cite{Berkovits:1997pq}
shares its internal structure with Berkovits' proposal for a nonpolynomial
superstring field theory. Its action and equation
of motion include two BRST-like generators \Gp and \Gtp (corresponding to the
BRST charge $Q$ and the field $\eta_0$ from the fermionization of the
world-sheet superghosts in the N=1 case), both mixing world-sheet bosons
and fermions (instead of matter and ghost fields for N=1 strings). The
main advantage of this model is its simplicity; no ghosts are needed in
addition to the matter fields. The equality of the structure and the
simplicity of the field realization of the BRST-like operators turn this
theory into a viable candidate for studying the intricacies which general
solutions to the equation of motion for nonpolynomial string field theory
bring about. Clearly, this equation of motion contains the star product.
Whereas the operator formulation~\cite{Gross:1986ia,Ohta:wn} and the 
diagonalization~\cite{Bars:2001ag,Douglas:2002jm,Belov:2002fp} 
of the bosonic vertex may be transferred literally to N=2 strings,
the fermionic part differs from the N=1 case since (after twisting, 
see section~\ref{sec:twist}) the fermions here have conformal weights 
0 and 1, respectively. Therefore, we commence the investigation of the 
operator formulation of the fermionic part of the star product in N=2 
string field theory, which is lacking in the present literature. This also has 
applications to Berkovits' hybrid formalism~\cite{Berkovits:1995ab}, where the 
compactification part of the theory is described by a twisted N=2 
superconformal algebra. Moreover, 
this $(0,1)$ first order system is isomorphic to the $(\xi,\h)$ ghost system 
of N=1 strings. Both, Witten's superstring field theory as well as 
Berkovits' nonpolynomial string field theory involve this ghost system in 
a nontrivial manner; the former features insertions of picture changing 
operators while the latter is formulated in the large Hilbert space. 
Therefore, a solid understanding of the structure of the star product in this 
sector is of interest and can easily be gleaned from the results presented 
here. Eventually, we point out that this fermionic system is also 
equivalent to the twisted $(b,c)$ system (cf.~end of section~\ref{sec:twist}). 
In the context of bosonic vacuum string field theory this auxiliary boundary 
conformal field theory was used to find solutions to the ghost equations of 
motion from surface state projectors constructed in the twisted $(b,c)$ 
system~\cite{Gaiotto:2001ji}. This BCFT is obtained by twisting the energy 
momentum tensor with the derivative of the ghost number current. The ghost 
fields of the twisted theory have conformal weights $(1,0)$ and thus 
correspond to the fermionic first order system of N=2 string field theory. 
However, it is unclear to us whether this equivalence can be traced to any 
deeper interrelation.  

In this paper we construct the vertices needed to formulate N=2 string 
field theory in the twisted fermionic sector from scratch using the operator 
language. In particular we pay attention to the anomaly of the $U(1)$ 
current $J$ contained in the \N2 superconformal algebra. Together with the 
overlap equations for the zero-modes this fixes the choice of vacuum for 
$N$-string vertices when one avoids midpoint insertions. For the identity 
vertex and the reflector the construction is accomplished using $\de$-function 
overlap conditions. The reflector is shown to implement BPZ conjugation as 
a graded antihomomorphism. To obtain the explicit form of the interaction 
vertex we have to invoke the Neumann function method. Supplemented with the 
above mentioned conditions on the vacuum the vertex is fixed. The Neumann 
coefficients are expressed in terms of coefficients of generating functions. 
We find an intimate relationship between the coefficients for the fermions 
and those for bosons allowing us to employ known identities from the 
boson Neumann matrices. Resorting to this relationship we show that the 
contribution of the $(0,1)$ system to the reparametrization anomaly cancels 
those of two real bosons. This is in accordance with their contribution 
$c=-2$ to the central charge. Finally, we explicitly check that the overlap 
equations for the interaction vertex are fulfilled. 

The paper is organized as follows. In the next section we briefly review the 
nonpolynomial string field theory for N=2 strings. The embedding of the N=2 
into a small N=4 superconformal algebra is described and the representation 
of the generators in terms of N=2 fields are given. Eventually, this 
small N=4 superconformal algebra is twisted, leading to a topological 
theory~\cite{Berkovits:1994vy}. After twisting all fields have integral 
conformal weights. In section~\ref{sec:id} the identity vertex is 
constructed. As a starting point $\de$-function overlap conditions for 
arbitrary $N$-string vertices are considered. After deducing the form of the 
identity vertex from the corresponding overlap equations its symmetries are 
discussed in detail with particular emphasis on the anomaly of the 
$U(1)$ current. In section~\ref{sec:refl} the 2-string vertex is considered. 
Starting from general $N$-string overlap equations formulated in terms of 
$\Z_N$-Fourier-transformed fields we discuss constraints on the vacua 
arising from the zero-mode overlap conditions. Avoiding midpoint insertions 
these conditions fix the vacuum on which the $N$-string vertex is built. 
The reflector is discussed as an application of the tools described in this 
section. A detailed discussion of BPZ conjugation as implemented by the 
2-vertex completes this section. We define BPZ conjugation and its 
inverse via the the bra-reflector and the ket-reflector, respectively, and 
their compatibility is shown. The interaction vertex is constructed in 
section~\ref{sec:intvert}. The Neumann coefficients of the $3$-string vertex 
are expressed in terms of generating functions constructed out of the 
conformal transformations which map unit upper half-disks into the 
scattering geometry of the vertex. The so obtained Neumann matrices are 
shown to be closely related to the Neumann matrices for bosons. Therefore, 
identities for the bosonic Neumann matrices entail corresponding identities 
for the fermionic ones. In this way, the anomaly of midpoint preserving 
reparametrizations is shown to cancel the contribution of two real bosons, 
which is in agreement with conformal field theory arguments. Finally, the 
overlap conditions for the interaction vertex are checked explicitly. Parts 
of this calculation are relegated to the appendix where also formulas for the 
bosonic vertices and Neumann coefficients are collected. The paper is 
concluded with a short summary and a discussion of possible 
applications and further developments. 

\section{Twisting the world-sheet action} \label{sec:twist}
\noindent
{\bf Nonpolynomial string field theory for N=2 strings.} The nonpolynomial
string field theory action~\cite{Berkovits:1997pq} coincides for N=1
(Neveu-Schwarz) and N=2 strings and contains in its Taylor expansion a
kinetic term and a cubic interaction similar to Witten's superstring field
theory action. In a notation suitable for both cases, it reads
\begin{equation}
S_{\text{SFT}}=\frac{1}{2g^2}\,\int\tr\left\{ 
     (e^{-\P}\Gp e^\P)(e^{-\P}\Gtp e^\P)
    -\int_0^1 dt (e^{-\Phh}\pa_t e^\Phh) \{ e^{-\Phh}\Gp e^\Phh,
    e^{-\Phh}\Gtp e^\Phh \} \right\} \, ; \label{eq:SFTact}
\end{equation}
here, $e^\P={\cal I}+\P+\frac{1}{2}\P\star\P+\ldots$ is defined via
Witten's midpoint gluing prescription (for convenience, all star products
in this action are omitted; ${\cal I}$ denotes the identity string field)
and $\P$ is a string field carrying $u(n)$ Chan-Paton labels with an
extension $\Phh(t)$ interpolating between $\Phh(t=0)=0$ and $\Phh(t=1)=
\P$. The BRST-like currents \Gp and \Gtp are the two superpartners of the
energy-momentum tensor in a twisted small N=4 superconformal algebra with
positive $U(1)$-charge. The action of \Gp and \Gtp on any string field is
defined in conformal field theory language as taking the contour integral,
e.\,g.\
\begin{equation}
  (\Gp e^\P)(z) = \oint \frac{dw}{2\pi i}\, \Gp(w) e^\P(z)\, , \quad
    (\Gtp e^\P)(z) = \oint \frac{dw}{2\pi i}\, \Gtp(w) e^\P(z)\, ,
\end{equation}
with the integration contour running around $z$. The corresponding
equation of motion reads
\begin{equation}
  \Gtp(e^{-\P} \star \Gp e^\P)=0 \, ,
\end{equation}
where contour integrations are implied again. We set out to define
unambiguously the star product and the integration symbol in~%
(\ref{eq:SFTact}).

\noindent
{\bf Small N=4 superconformal algebra.} Just as in bosonic and in superstring 
theory, the critical dimension for N=2 string theory can be determined
from anomaly considerations. It turns out that such a theory has
propagating on-shell degrees of freedom only in signature (2,2); thus,
the (K\"{a}hler) spacetime is naturally parametrized by two complex
bosons $Z^a$,
\begin{equation}
  Z^0 := X^1 + \ic X^2\, , \quad Z^1 := X^3 + \ic X^4 \, .
\end{equation}
Their complex conjugates are denoted by $\Zb^\ab, \ab\in\{\nb,\ob\}$.
The metric on flat $\C^{1,1}$ is taken to be $(\eta_{a\ab})$ with
nonvanishing components $\eta_{1\bar{1}}=-\eta_{0\bar{0}}=1$. To
obtain N=2 world-sheet supersymmetry, the four real bosons have to
be supplemented by four Dirac spinors $\ps^\mu$ which may be combined
into
\begin{equation}
  \ps^{+0} := \ps^1 + \ic \ps^2\, , \quad \ps^{+1} := \ps^3 + \ic
  \ps^4\, , \quad \ps^{-\nb} := \ps^1 - \ic \ps^2\, , \quad \ps^{-\ob}
  := \ps^3 - \ic \ps^4 \, . \label{eq:psidef}
\end{equation}
For open strings, we apply the doubling trick throughout this paper
so that all fields are (holomorphically) defined on the double
cover of the disk, i.e., on the sphere. In superconformal gauge, the
world-sheet action for the matter fields now reads (on a Euclidean
world-sheet $\S$ with double cover~$\tilde{\S}$)
\begin{equation}
  S = \frac{1}{4\pi\a'}\int_{\tilde{\S}} dz\wedge d\zb\, (\pa Z\cdot
    \pab\Zb + \pab Z\cdot\pa\Zb) + \frac{1}{8\pi} \int_{\tilde{\S}}
    dz\wedge d\zb\, (\ps^+ \cdot\pab\ps^- + \ps^-\cdot\pab\ps^+)\, .
  \label{eq:untwist}
\end{equation}
The action is normalized in such a way that the operator product
expansions are the ones which should be expected from the transition
from real to complex coordinates:
\begin{equation}
  Z^a(z) \Zb^\ab(w)\sim -\a'\eta^{a\ab}\ln|z-w|^2\, , \quad
  \ps^{+a}(z)\ps^{-\ab}(w)\sim\frac{2\eta^{a\ab}}{z-w}\, .\label{eq:OPEs}
\end{equation}
The matter part of the constraint algebra for this theory is an N=2
superconformal algebra with generators
\begin{gather}
  T = -\frac{1}{\a'}\pa Z\cdot\pa\Zb - \frac{1}{4}(\ps^+ \cdot\pa\ps^-
    + \ps^-\cdot\pa\ps^+)\, ,\notag \\
  \Gp = \frac{\ic}{\sqrt{2\a'}}\ps^+\cdot \pa\Zb\, ,\quad
  \Gm = \frac{\ic}{\sqrt{2\a'}}\ps^-\cdot \pa Z\, , \label{eq:N2SCA} \\
  J = \frac{1}{2} \ps^+\cdot\ps^-\, . \notag
\end{gather}
In $D=4$, the central charge is $c=6$ (as required from the ghosts,
which we will, however, not introduce). Note that the superscripts $\pm$
on each quantity label the charge under the $U(1)$ current $J$. These
currents can, in principle, be defined on general $D$-dimensional K\"ahler
manifolds for any $D\in 2\Na$.

In $D=4$, we can extend the N=2 superconformal into a small N=4
superconformal algebra with additional generators\footnote{On a $D$-%
dimensional manifold, $\ve_{ab}$ and $\ve_{\ab\bb}$ have to be replaced
by (the components of) nondegenerate $(2,0)$- and $(0,2)$-forms,
respectively. (Untwisted) N=4 supersymmetry requires a hyperk\"ahler
spacetime manifold.}~\cite{Lechtenfeld:2002cu}
\begin{gather}
\begin{split}
  J^{++} = \frac{1}{4}\ve_{ab}\,\ps^{+a}\ps^{+b}\, , \quad
  J^{--} = \frac{1}{4}\ve_{\ab\bb}\,\ps^{-\ab}\ps^{-\bb}\, , \\
  \Gtp = \frac{\ic}{\sqrt{2\a'}} \ve_{ab}\, \ps^{+a}\pa Z^b\, , \quad
  \Gtm = \frac{\ic}{\sqrt{2\a'}} \ve_{\ab\bb}\, \ps^{-\ab}\pa \Zb^\bb\, ,
  \label{eq:N4SCA}
\end{split}
\end{gather}
using the constant antisymmetric tensor $\ve$ with $\ve_{01}=\ve_{\nb\ob}=
-\ve^{01}=-\ve^{\nb\ob}=1$. The currents $J$, $J^{++}$ and $J^{--}$ form
an affine $su(1,1)$ Kac-Moody algebra of level 2.

In order to obtain an algebra with central charge zero, we can twist the
small N=4 superconformal algebra by shifting $T\to T':=T+\frac{1}{2}\pa
J$, i.e., {\em reducing} the weight of a field by one half of its charge.
After twisting, all fields will have integral weights; in particular,
\Gp and \Gtp as fields of spin~1 may subsequently serve as BRST-like
currents.  We will show later that the bosonic and fermionic contributions
to the anomalies of all currents in the small N=4 superconformal algebra
on $N$-vertices cancel for all even~$D$. Of course, the definition of
$\ve_{ab}$ requires at least two complex dimensions, i.\,e., $D\geq 4$.

\noindent
{\bf Twisted action.} With respect to the twisted energy-momentum tensor
$T'$, $\ps^{+a}$ and $\ps^{-\ab}$ have weights~0 and~1; this suggests that
they are no longer complex conjugates in the sense of eq.~(\ref{eq:psidef}).
Indeed, they constitute a first order system with Euclidean world-sheet
action
\begin{equation}
  S_\psi' = \frac{1}{4\pi} \int_{\tilde{\S}} dz\wedge d\zb\, \ps^+ \cdot
    \pab\ps^-
  \label{eq:twist}                                                        
\end{equation}                                                            
which is real after a Wick back-rotation to Minkowski space     
for {\em hermitean} fields $\ps^\pm$.\footnote{Here and in  
the following, we sometimes omit the spacetime labels on $\ps^\pm$ if the
statement refers to any of the $\ps^{+a}, \ps^{-\ab}$.} That this action for
the fermionic part of the twisted theory is indeed the correct one is
corroborated by the fact that the full action is invariant under the
symmetries generated by all currents in the small twisted N=4 superconformal
algebra.

As fields of integral weight, both $\ps^+$ and $\ps^-$ are integer-moded.
In particular, the spin~0 field $\ps^+$ has a zero-mode on the sphere. In
analogy to the $bc$-system there are thus two vacua at the same energy
level: the bosonic $SL(2,\R)$-invariant vacuum $|0\>=:\dv$ is annihilated
by the Virasoro modes $L_{m\geq -1}$ and $\ps^+_{m>0}$, $\ps^-_{m\geq 0}$;
its fermionic partner, $\uv:=\ps^+_0\dv$, is annihilated by $\ps^+_{m\geq
0}$, $\ps^-_{m>0}$. To get nonvanishing fermionic correlation functions, we
need one $\ps^+$-insertion, i.e., $\<\downarrow\!\dv = \<\uparrow\!\uv = 0$,
$\<\downarrow\!\uv = 1$. Taking into account the odd background charge
we assign to {\it dual} vacua (e.g., $\uv$ and $\dvb$) the same Grassmannality.

\noindent
{\bf Translation to the $\boldsymbol{\eta\xi}$ and twisted 
$\boldsymbol{bc}$ system.} All methods to construct string field theory 
vertices used in this paper can be formulated in terms of conformal field 
theory data. In particular, they only depend on the conformal weights and 
the world-sheet statistics of the fields. The $\ps^-\ps^+$ system as 
well as the $\eta\xi$ system from the fermionization of the world-sheet 
superghosts and the twisted $bc$ system of~\cite{Gaiotto:2001ji} are 
fermionic first order systems with fields of conformal weights~1 and~0. 
Therefore, all formulas in this paper remain valid upon the 
substitutions\footnote{These substitutions can be used to compare part of our 
results on the 3-vertex with those obtained in~\cite{Maccaferri:2003rz}, a 
preprint which appeared on the same day.} 
\begin{subequations}
\begin{gather}
\ps^+\quad\longleftrightarrow\quad\sqrt{2}\,c'
\quad\longleftrightarrow\quad\sqrt{2}\,\xi\,, \\ 
\ps^-\quad\longleftrightarrow\quad\sqrt{2}\,b'
\quad\longleftrightarrow\quad\sqrt{2}\,\eta\,.
\end{gather}
\end{subequations}
The factors of $\sqrt{2}$ take care of the unusual normalization of the 
$\ps^-\ps^+$ two-point function, which can be read off from 
eq.~(\ref{eq:OPEs}). For example, eq.~(\ref{eq:twist}) translates via the 
dictionary given above into the action for the $\eta\xi$ system . 
Furthermore, states built from $\dv$ are in the small Hilbert space whereas 
states constructed from $\uv$ are in the large Hilbert space 
of~\cite{Friedan:1985ge}. 

\section{Identity vertex} \label{sec:id}
\noindent
The identity vertex defines the integration in eq.~(\ref{eq:SFTact}); it
is an element $|\Ical\>$ of the one-string Hilbert space corresponding to
the identity string field $\Ical$. The identity vertex glues the left and
right halves of a string together; therefore it can be defined via the
corresponding overlap equations.

\noindent
{\bf Overlap equations.} In general, the overlap equations for an $N$-vertex
can be determined from conformal field theory arguments~\cite{Gross:1987pp}:
On the world-sheet of the $r$-th string ($r\in\{1,\ldots,N\}$), a strip, we
introduce coordinates $\xi_r=\t_r+\ic\s_r$. The strip can be mapped into an
upper half-disk with coordinates $z_r=e^{\xi_r}$; the upper half-disks are then
glued together in the scattering geometry in a such a way that
\begin{equation}
  z_r z_{r-1} = -1 \qquad\text{for }|z_r| = 1\, , \:\: \text{Re}(z_r)\geq 0\, ,
  \quad\text{i.\,e.,}\quad 0\leq\s_r\leq\frac{\pi}{2}\, ,\;\; \t_r = 0\, . 
\label{eq:crdglu}
\end{equation}
This is achieved by the conformal map
\begin{equation}
  f_r(z) = - e^{\ic\pi\frac{1-2r}{N}} f(z)\, , \qquad f(z) = \left(
    \frac{1+\ic z}{1-\ic z}\right)^{2/N}\, , \label{eq:fr}
\end{equation}
where the phases have been chosen so as to give a symmetric configuration
when mapping back to the upper half-plane.

A primary field $\phi^{(r)}$ of conformal weight~$h$ in the boundary
conformal field theory on the strip is glued according to
\begin{equation} \label{eq:gluing}
\begin{split}
  \phi^{(r)}(\s_r,\t_r=0) & \equiv \phi^{(r)}(\xi_r)
    = z_r^h \phi^{(r)}(z_r) \\
    & = \left( z_r\frac{\pa z_{r-1}}{\pa z_r}\right)^h \phi^{(r-1)}(z_{r-1})
    = \left(\frac{z_r}{z_{r-1}}\frac{\pa z_{r-1}}{\pa z_r}\right)^h
    \phi^{(r-1)}(\s_{r-1},\t_{r-1}=0) \\
    & = (-1)^h \phi^{(r-1)}(\pi-\s_{r},\t_r=0) \qquad\text{for }
    0\leq\s_r\leq\frac{\pi}{2}\, .
\end{split}
\end{equation}
In the last two lines we have used~(\ref{eq:crdglu}). This equality is
required to hold when applied to the $N$-string vertex $\<V_N|$. 
If we insert the open string mode expansion for $\t=0$, 
$\phi^{(r)}(\s)=\phi^{(r)}_0+\sum_n(\phi^{(r)}_n+\phi^{(r)}_{-n})\cos n\s$, 
we obtain a condition on the modes. For
$N\leq 2$, the above condition extends to $0\leq\s\leq\pi$, so that
one can take advantage of the orthogonality of the cosine to obtain the
diagonal condition
$\<V_N|\big( \phi^{(r)}_n+\phi^{(r)}_{-n}+(-1)^{n+h}(\phi^{(r-1)}_{-n}+
\phi^{(r-1)}_n)\big)=0$. Instead, we will impose the stricter
condition $\<V_N|\big( \phi^{(r)}_n+(-1)^{n+h}\phi^{(r-1)}_{-n})\big)=0$.
For $N>2$, the overlap equations in general mix all modes.

\noindent
{\bf Construction of the identity vertex.} For the $\ps^\pm$-system,
we demand the stricter conditions
\begin{subequations} \label{eq:idov}
\begin{align}
  \<\Ical|\big[ \ps^+_n -(-1)^n \psi^+_{-n} \big]  &= 0 & \Longrightarrow
    & & \<\Ical|\ps^+(\s) & = \<\Ical|\ps^+(\pi-\s) \, ,\label{eq:idovp} \\
  \<\Ical|\big[ \ps^-_n +(-1)^n \psi^-_{-n} \big] &= 0 & \Longrightarrow
    & & \<\Ical|\ps^-(\s) & = -\<\Ical|\ps^-(\pi-\s)\, ,
\end{align}
\end{subequations}
from which the gluing conditions~(\ref{eq:gluing}) follow.
The conditions on $\<\Ical|$ are compatible since $\{\ps^+_n -(-1)^n
\psi^+_{-n}, \ps^-_n +(-1)^n \psi^-_{-n}\}=0$. The obvious solution
to eqs.~(\ref{eq:idov}) reads
\begin{equation}
\begin{split}
  \<\Ical| & = \dvb\prod_{n=1}^\infty \frac{1}{2}\big[ \ps^+_n -(-1)^n
    \psi^+_{-n} \big] \big[ (-1)^n \ps^-_n +\psi^-_{-n} \big] \\
    & = \dvb\prod_{n=1}^\infty \big[ 1+\frac{1}{2}(-1)^n \ps^+_{n}
    \psi^-_{n} \big] = \dvb\exp\Big[ \frac{1}{2}\sum_{n=1}^\infty
    (-1)^n \ps^+_{n} \psi^-_{n} \Big] \, , \label{eq:idver}
\end{split}
\end{equation}
where the $SL(2,\R)$-invariant vacuum $\dvb$ is annihilated by $\ps^-_0$.

\noindent
{\bf Symmetries of the vertex.} Applying the gluing conditions~%
(\ref{eq:gluing}) to the complex spin~1 fields
\begin{equation}
  \pa Z = -\ic\sqrt{\frac{\a'}{2}}\sum_k \a_k z^{-k-1}\qquad\text{and}\qquad
  \pa\Zb = -\ic\sqrt{\frac{\a'}{2}}\sum_k \bar{\a}_k z^{-k-1}\, ,
\end{equation}
we obtain
\begin{equation}
  \<\Ical|(\a_n+(-1)^n \a_{-n}) = 0 \, , \qquad \<\Ical|(\bar{\a}_n+(-1)^n
    \bar{\a}_{-n}) = 0\, . \label{eq:idovbos}
\end{equation}
Together with~(\ref{eq:idov}), this entails that the gluing conditions
for the BRST-like spin~1 currents \Gp and $\Gtp$,
\begin{equation}
 \<\Ical|(G^+_n + (-1)^n G^+_{-n}) = 0\, , \qquad
  \<\Ical|(\tilde{G}^+_n + (-1)^n \tilde{G}^+_{-n}) = 0\, ,
\end{equation}
are satisfied. In general, anomalies can only appear if the current
contains pairs of conjugate oscillators. Thus, it is
clear that the spin~2 currents $J^{--}$, \Gm and \Gtm are anomaly-free,
just like the spin~0 current $J^{++}$. More interesting are the (twisted)
energy-momentum tensor and the $U(1)$ current~$J$ (when treated as
primary fields).

The modes of the twisted energy-momentum tensor
$T' = -\frac{1}{\a'}\pa Z\cdot\pa\Zb-\frac{1}{2}\ps^-\cdot\pa\ps^+$
can be written as 
\begin{equation}
  L_n = \frac{1}{2}\sum_m \a_m\cdot \bar{\a}_{n-m} + \frac{1}{2}\sum_m
    (n-m)\ps^-_m\cdot\ps^+_{n-m} \, .
\end{equation}
According to~(\ref{eq:gluing}) these modes have to satisfy\footnote{
Treating the energy-momentum tensor as a primary field is justified iff 
the central
charge vanishes. In this sense, one can understand eq.~(\ref{eq:idKnAn}) as
a condition on the central charge. Note that the $N$-string variant
$\<V_N|\sum_{r=1}^N (L^{(r)}_n-(-1)^n L^{(r)}_{-n})=0$ of eq.~(\ref{eq:idKnAn})
does not follow from eq.~(\ref{eq:gluing}) for $N>2$.}
\begin{equation}
  \<\Ical|K_n := \<\Ical|(L_n-(-1)^n L_{-n}) = 0 \label{eq:idKnAn}
\end{equation}
for the vertex to be reparametrization invariant. In $D/2$ complex
dimensions, the contribution of the bosons to the left hand side of eq.~%
(\ref{eq:idKnAn}) can be easily shown to be
\begin{equation}
  \<\Ical|K_{2n}^\a = \frac{D}{2}(-1)^{n} n\,\<\Ical|\, ,
\end{equation}
which is canceled by the fermionic contribution
\begin{equation}
  \<\Ical|K_{2n}^\ps = -\frac{D}{2} (-1)^n n\,\<\Ical| \, .
\end{equation}
These contributions arise from terms $\frac{1}{2}\a_n\cdot\bar{\a}_n$
and $\frac{n}{2}\ps^-_n\cdot\ps^+_n$ in $K_{2n}^\a$ and $K_{2n}^\ps$,
respectively. Due to the absence of such terms, the $K_{2n+1}$ 
are automatically anomaly-free.

Before considering the $U(1)$ current $J$, let us first recall the 
discussion in~\cite{Witten:1985cc} of the $U(1)$-anomaly of $N$-vertices:
If the current $J$ is bosonized as $J=\pa\vp$, the action for
this boson reads
\begin{equation}
  S = -\frac{1}{4\pi}\int dz\wedge d\zb\, (\pa\vp\bar{\pa}\vp +
    Q R\vp)\, .
\end{equation}
The operator product expansion is that of the free action, 
$\vp(z)\vp(w)\sim\ln(z-w)$. The energy-momentum tensor for $\vp$ reads 
$T_\vp=\frac{1}{2} J^2-Q\pa J$, where $Q$ is the background charge, i.\,e., 
the coefficient of the third order pole in the operator product expansion
$T(z)J(w)$. For the $\ps^+ \ps^-$-system in $D/2$ complex dimensions,
$Q=-D/2$.

In a general gluing geometry the curvature is concentrated in one point, 
namely the
midpoint of the string ($\s=\pi/2$). On such surfaces the term linear 
in $\vp$ contributes an anomalous factor of
\begin{equation}
  \exp\Big( \frac{Q}{2\pi}\vp(\pi/2)\int d^2\s R\Big)
    \label{eq:anomPI}
\end{equation}
in the path integral\footnote{In the integral of the Ricci scalar over this 
surface we have used $dz\wedge d\zb=2d\s d\t$.}. 
This integral measures the deficit angle of
this surface when circumnavigating the curvature singularity at the string
midpoint and contributes $-(N-2)\pi$ for an $N$-string 
vertex. Hence, the factor~(\ref{eq:anomPI}) produces a $U(1)$-anomaly of 
$(N-2)\frac{D}{4}$ in the path integral. Since the $U(1)$-charge\footnote{%
The $U(1)$ charge of a bra vector is measured by $J_0^{\dagger}$; we use
conventions where $J_0=\sfrac{1}{2}\sum_{m=1}^{\infty}(\psi^+_{-m}\cdot
\psi^-_{m}-\psi^-_{-m}\cdot\psi^+_{m})+\sfrac{1}{2}\psi^+_{0}\cdot\psi^-_{0}
-\sfrac{D}{4}=-J_0^{\dagger}$.} of ${\dvb}$ is $-\frac{D}{4}$, an $N$-vertex
constructed from $N$ ${\dvb}$-vacua requires $(N-2)\frac{D}{4}-N\left(
-\frac{D}{4}\right)=\frac{D}{2}(N-1)$ $\ps^+$-insertions (the exponential
factor is neutral).\footnote{Alternatively, we could use midpoint insertions
to adjust the anomaly of the vertex. However, in our cases, they make life
unnecessarily complicated. Since the value of the anomaly,
$(N-2)\frac{D}{4}$, never exceeds the maximal charge of $\<\Om_N|$,
i.\,e. $N\frac{D}{4}$ for $\<\Om_N|_\text{max}=\uvb^{\otimes N}$, we
can avoid midpoint insertions.} This is consistent with~(\ref{eq:idver})
for $N=1$.

Therefore, we do not expect the $U(1)$-current $J$ to be anomaly-free;
\begin{equation}
  \<\Ical|(J_n+(-1)^n J_{-n}) \neq 0 \label{eq:idverjan}
\end{equation}
in general. Since its zero-mode measures the fermion number of the
vertex, we instead expect $\<\Ical|J_0 = \frac{D}{4}\<\Ical|$. This
relation holds trivially. For $n\neq 0$ in eq.~(\ref{eq:idverjan}),
one obtains $\<\Ical|(J_{2n}+J_{-2n})=(-1)^{n}\frac{D}{2}\<\Ical|$.

\section{Reflector}\label{sec:refl}
\noindent
In this section we construct the 2-string vertex for the fermionic
$(1,0)$ system $(\ps^-,\ps^+)$. It is convenient to introduce
$\mathbbm{Z}_N$-Fourier-transformed fields as a tool to diagonalize
general $N$-string overlap equations. The overlap equations fix the zero-mode
part of the 2-vertex up to a sign. A discussion of BPZ conjugation
motivates our choice for this sign.

\noindent
{\bf $\boldsymbol{\mathbbm{Z}_N}$-transforms.} Introducing the combinations
\begin{equation}
\sum_n\psi^-_n\,e^{\pm\ic n\s}=\pi_{\psi^+}(\s)\pm\ic\,\psi^-(\s)\,, \qquad 
\sum_n\psi^+_n\,e^{\pm\ic n\s}=\psi^+(\s)\pm\ic\,\pi_{\psi^-}(\s) 
\end{equation}
of left and right movers, the conditions imposed on the fermions following 
from the $\de$-function overlap of $N$ strings are
\begin{subequations}
\label{eq:Novlap}
\begin{align}
  \ps^{+(r)}(\s) & = \begin{cases}
            \ps^{+(r-1)}(\pi-\s),\quad\s\in[\,0\,,\sfrac{\pi}{2}]\, , \\
            \ps^{+(r+1)}(\pi-\s),\quad\s\in[\sfrac{\pi}{2},\pi]\, ,
                   \end{cases} \\
  \pi^{(r)}_{\psi^+}(\s) & = \begin{cases}
          -\pi^{(r-1)}_{\psi^+}(\pi-\s),\quad\s\in[\,0\,,\sfrac{\pi}{2}]\, , \\
          -\pi^{(r+1)}_{\psi^+}(\pi-\s),\quad\s\in[\sfrac{\pi}{2},\pi]\, .
                   \end{cases}
\end{align}
\end{subequations}
For $\psi^-(\s)$ and $\pi_{\psi^-}(\s)$ similar equations have to be fulfilled.
The conditions (\ref{eq:Novlap}) are easily diagonalized if we introduce
$\mathbbm{Z}_N$-Fourier-transformed fields~\cite{Gross:1986ia},
\begin{subequations} \label{eq:Z2}
\begin{align}
  \Psi^a(\s) & = \frac{1}{\sqrt{N}}\sum_{r=1}^{N}\ps^{+(r)}(\s)
    e^{\frac{2\pi\ic ra}{N}}\, , &
  \Pi^a(\s) & = \frac{1}{\sqrt{N}}\sum_{r=1}^{N}\pi^{(r)}_{\psi^+}(\s)
    e^{\frac{2\pi\ic ra}{N}}\, ,
\end{align}
\end{subequations}
where $a\in\{1,\dots,N\}$. 
Note that now $(\Psi^a, \Pi^{N-a})$ form canonically conjugate
pairs (the upper index is taken modulo $N$). We choose the following
ansatz for the $N$-vertex in terms of $\mathbbm{Z}_N$-transformed
oscillators
\begin{equation}
  \<V_N| =  \<\Om_N|\exp\big( \sfrac{1}{2}\sum_{a=1}^N\sum_{m,n} \Psi_{m}^a
    V^a_{mn} \Pi_{n}^{N-a} \big) \label{eq:VNZN}
\end{equation}
with Neumann matrices $V^a$ and a vacuum state $\<\Om_N|$.
The vacuum state will be determined below from the zero-mode overlap
conditions; the summation range of $m,n$ should then be adjusted in such
a way that only creation operators w.\,r.\,t.\ this vacuum appear in the
exponential.

In application to $\<V_N|$, eqs.~(\ref{eq:Novlap}) now read
\begin{subequations}
\label{eq:ZNov}
\begin{align}
  \<V_N|\Psi^a(\s) & = \begin{cases}
                   e^{\frac{2\pi\ic a}{N}}\<V_N|\Psi^a(\pi-\s)\, ,\\
                   e^{\frac{-2\pi\ic a}{N}}\<V_N|\Psi^a(\pi-\s)\, ,
                       \end{cases} \\
  \<V_N|\Pi^a(\s)  & = \begin{cases}
                   -e^{\frac{2\pi\ic a}{N}}\<V_N|\Pi^a(\pi-\s)\, ,\\
                   -e^{\frac{-2\pi\ic a}{N}}\<V_N|\Pi^a(\pi-\s)\, .
                      \end{cases}
\end{align}
\end{subequations}
As already discussed in section~\ref{sec:id}, the overlap conditions
will only contain a sum of two oscillators (rather than infinitely many),
if after inserting the mode expansions the cosines can be integrated
over $[\,0,\pi]$. This is obviously possible also for $N>2$ if
$\frac{2a}{N}\in\mathbbm{N}$. Therefore, $(\Psi^N,\Pi^N)$ and, if $N$
is even, $(\Psi^{N/2},\Pi^{N/2})$ appear in the vertex~(\ref{eq:VNZN})
with Neumann matrices $V^N=-C$ and $V^{N/2} = C$, respectively.
Here, $C$ denotes the twist matrix with components $C_{mn}=(-1)^m\de_{mn}$.

Before we turn to the 2-string vertex, let us briefly discuss the
overlap conditions for the zero-modes of the $\mathbbm{Z}_N$-transformed
oscillators. It is consistent with~(\ref{eq:ZNov}) to demand
\begin{subequations}
\label{eq:VNvac}
\begin{align}
  \<\Om_N|\Psi^a_0 & = 0 \qquad\text{for }1\leq a\leq N-1\, ,
    \label{eq:VNPsia0} \\
  \<\Om_N|\Pi^N_0 & = 0\, . \label{eq:VNPiN0}
\end{align}
\end{subequations}
Note that eqs.~(\ref{eq:VNPsia0}) entail that no $\Psi^a_0$ (for
$a\in\{1,\ldots,N-1\}$) may appear in the exponential of the vertex~%
(\ref{eq:VNZN}). The appearance of $\Psi^N_0$ is forbidden by eq.~%
(\ref{eq:VNPiN0}) since $V^N=-C$ is diagonal. In terms of the original
one-string oscillators, this means that no $\psi^+_0$ occurs in the
exponential of the vertex.

It is easy to see that the conditions on the vacuum~(\ref{eq:VNvac}) are
solved by\footnote{Here one has to use the fact that $\uvb$ is 
Grassmann {\it{even}} while $\dvb$ is Grassmann {\it{odd}}, i.e., the 
bra-vacua have opposite
Grassmannality compared to the corresponding ket-vacua.
This is a consequence of the odd background charge (cf.\ the end of
section~\ref{sec:twist}).}
\begin{equation}
  \<\Om_N| = \pm\sum_{k=1}^N \stl{1}\uvb\otimes\ldots\stl{k-1}\uvb
    \otimes\stl{k}\dvb\otimes\stl{k+1}\uvb\otimes\ldots\stl{N}\uvb \, .
    \label{eq:Nvac}
\end{equation}
The subscripts indicate in which string Hilbert space the corresponding
vacuum state lives. The vacuum (\ref{eq:Nvac}) already features the $U(1)$ 
charge required by the $J$-anomaly, namely $(N-2)\frac{D}{4}$. This
choice allows us to avoid midpoint insertions.

\noindent
{\bf Overlap equations for the reflector.} Expressed in terms of
$\Z_2$-transforms, the overlap conditions for the reflector simply become
\begin{subequations}
\label{eq:ovlapV2Z2}
\begin{align}
  \<V_2|\Psi^1(\s) & = -\<V_2|\Psi^1(\pi-\s) \, , &
  \<V_2|\Psi^2(\s) & = \<V_2|\Psi^2(\pi-\s) \, , \\
  \<V_2|\Pi^2(\s) & = -\<V_2|\Pi^2(\pi-\s) \, , &
  \<V_2|\Pi^1(\s) & = \<V_2|\Pi^1(\pi-\s)\, ,
\end{align}
\end{subequations}
which can be rewritten in terms of modes acting on $\<V_2|$ as
\begin{subequations}
\label{eq:modV2Z2}
\begin{align}
  (\Psi^1_m +\Psi^1_{-m}) & = -(-1)^m(\Psi^1_m +\Psi^1_{-m})\,,
  &(\Psi^2_m +\Psi^2_{-m})&=(-1)^m(\Psi^2_m +\Psi^2_{-m})\, ,\\
  (\Pi^1_m + \Pi^1_{-m}) &=(-1)^m(\Pi^1_m + \Pi^1_{-m})\,,
  &(\Pi^2_m + \Pi^2_{-m})&=-(-1)^m(\Pi^2_m + \Pi^2_{-m})\,,
\end{align}
\end{subequations}
for the nonzero-modes. The conditions for the zero-modes read
\begin{equation}
\label{eq:V2zerom}
  \<V_2|\Psi^1_0 = 0\, , \qquad \<V_2|\Pi^2_0 = 0 \, .
\end{equation}
The zero-modes $\Psi^2_0$ and $\Pi^1_0$ put no restrictions on the vertex.
Along the lines of~\cite{Gross:1986ia}, one finds  
\begin{subequations}\label{eq:V2bra}
\begin{align}
  \<V_2| & = \<\Om_2| \exp\Big( \sfrac{1}{2}\sum_{m=1}^{\infty}\big[
    \Psi^2_{m} (-1)^m\Pi^2_{m}-\Psi^1_{m}(-1)^m\Pi^1_{m}\big] \Big) \\
    & = \<\Om_2| \exp\Big( \sfrac{1}{2}\sum_{m=1}^{\infty}\big[
    \psi^{+(1)}_m (-1)^m\psi^{-(2)}_{m} + \psi^{+(2)}_{m} (-1)^m
    \psi^{-(1)}_m \big] \Big)
\end{align}
\end{subequations}
as a solution to eqs.~(\ref{eq:modV2Z2}). 
Since no zero-modes appear in the vertex, the vacuum $\<\Om_2|$ has to be
annihilated by $\Psi^1_0$ and $\Pi^2_0$ in order to satisfy 
eq.~(\ref{eq:V2zerom}). Thus the vacuum is a {\it symmetric} combination 
of up- and down-vacua in the two-string Hilbert space,
\begin{equation}
  \<\Om_2| = \pm(\stl{1}\uvb\otimes\stl{2}\dvb + \stl{1}\dvb\otimes\stl{2}\uvb)
    =: \pm(\udvb + \duvb).
\end{equation}
This is consistent with eq.~(\ref{eq:Nvac}). In the last expression it
is understood that the first entry corresponds to string $1$,
while the second corresponds to string $2$. The overall sign is
determined by requiring that $\<V_2|$ implements BPZ conjugation.

\noindent
{\bf BPZ conjugation.} On a single field $\phi(z)$ BPZ conjugation acts as
$I\circ\phi(z)$ with $I(z)=-1/z$; since $I$ inverts the time direction,
it is suggestive that on a product of fields, BPZ conjugation should
reverse the order of the fields. This statement will be put on a more solid 
ground below. The action of BPZ on fields induces an action on states:
$\text{bpz}(|\phi\>)$ defines the out-state $\<\phi|$ which is created
by $\lim_{z\to\infty}\<0|I\circ\phi(z)$. In terms of modes this prescription 
yields 
\begin{equation}
  \text{bpz}(\phi_n)=(-1)^{n+h}\phi_{-n} \,  \label{eq:bpz}
\end{equation} 
for a field of conformal weight $h$. 
To fix the choice of vacuum in (\ref{eq:V2bra}), recall
that $\<V_2|$ is an element of the tensor product of two dual string
Hilbert spaces $\<V_2|\in \Hcal^*\otimes\Hcal^*$ and thus induces an
odd linear map from $\Hcal$ to $\Hcal^*$, which is nothing but BPZ
conjugation~\cite{Gaberdiel:1997ia},
\begin{equation}
  \<V_2|\phi\>\stl{1}=\stl{2}\<\text{bpz}(\phi)| \, . \label{eq:bpzV2}
\end{equation}
In order to be compatible with the usual definitions of BPZ conjugation, 
we demand in particular 
that the $SL(2,\R)$ invariant vacuum $\dv$ is mapped into $\dvb$ under BPZ 
conjugation. Therefore we fix the vacuum $\<\Om_2|$ to be
\begin{equation}
  \<\Om_2| = \stl{1}\uvb\otimes\stl{2}\dvb + \stl{1}\dvb\otimes\stl{2}\uvb
    =: \udvb + \duvb\,.
\end{equation}
Note that with this choice of vacuum and using eq.~(\ref{eq:bpzV2}) one finds
$\bpz(\dv)=\dvb$ and $\bpz(\uv)=\uvb$. 

Now consider the corresponding ket state $|V_2\>$. Observe that the 
conformal transformation $I$ maps $(\t,\s)$ to $(-\t,\pi-\s)$. 
Therefore, the overlap equations for $0\le\s\le\pi/2$ for a field 
$\phi$ of conformal weight $h_{\phi}\,$, 
$\phi^{(r)}(\s)=(-1)^{h_{\phi}}\phi^{(r-1)}(\pi-\s)\,$,
transform into $\phi^{(r)}(\s)=(-1)^{-h_{\phi}}\phi^{(r+1)}(\pi-\s)$. 
This implies that the overlap equations for the $N=1$ and $N=2$ vertices are 
invariant under BPZ conjugation for fields of integral conformal weight. 
Indeed, this can be verified for~(\ref{eq:modV2Z2}) using~(\ref{eq:bpz})
on the level of modes, and we can immediately write down the solution
\begin{subequations}
\begin{align}
  |V_2\> & = \exp\Big( \sfrac{1}{2}\sum_{m=1}^{\infty}\big[ \Pi^2_{-m}
    (-1)^m \Psi^2_{-m}-\Pi^1_{-m}(-1)^m \Psi^1_{-m}\big] \Big)|\Om_2\> \\
    & = \exp\Big( \sfrac{1}{2}\sum_{m=1}^{\infty}\big[ \ps^{-(1)}_{-m}
    (-1)^m\psi^{+(2)}_{-m} + \ps^{-(2)}_{-m} (-1)^m\psi^{+(1)}_{-m} \big]
    \Big) |\Om_2\> \, . \label{eq:V2ket}
\end{align}
\end{subequations}
It is easy to see that eqs.~(\ref{eq:modV2Z2}), now taken to act on the ket
vertex, are fulfilled. Eventually we have to fix our choice of vacuum.
In order to fulfill the zero-mode overlap equations (\ref{eq:V2zerom}),
$|\Om_2\>$ has to be an {\it antisymmetric} combination of up and down vacua
\begin{equation} \label{eq:V2vac}
  |\Om_2\> = \pm\,(\uv\stl{1}\otimes\dv\stl{2} - \dv\stl{1}\otimes
    \uv\stl{2}) = \pm\,(\udv - \duv) \, .
\end{equation}
We fix the overall sign of $|\Om_2\>$ to be a plus sign by requiring
\begin{equation} \label{eq:bpzV2inv}
  \bpz^{-1}(\<\phi|) := \stl{2}\<\phi|V_2\>\stl{12} (-1)^{|\phi|+1}
    = |\bpz^{-1}(\phi)\>\stl{1} \, ,
\end{equation}
where $|\phi|$ denotes the Grassmannality of the state $|\phi\>$.
Moreover one finds
\begin{eqnarray}
\label{eq:grantihom}
  \bpz(\ps^+_{-k}\ps^-_{-l}\dv) & = &
    \stl{12}\<V_2|\ps^{+(1)}_{-k}\ps^{-(1)}_{-l}\dv\stl{1} =
  \stl{12}\<V_2|\ps^{+(2)}_k (-1)^k\ps^{-(1)}_{-l}\dv\stl{1} \nonumber \\
  & = & \stl{12}\<V_2|\ps^{-(1)}_{-l} \ps^{+(2)}_{k} (-1)^k (-1)^{|\psi^{+}|
    |\psi^{-}|} \dv\stl{1}  \\
  & = & \stl{2}\dvb\psi^{-(2)}_l \psi^{+(2)}_k (-1)^{k+l+1}
  (-1)^{|\psi^{+}||\psi^{-}|} \, ,\nonumber
\end{eqnarray}
which can be checked using~(\ref{eq:V2bra}). Eq.~(\ref{eq:grantihom})
is the statement that BPZ conjugation acts as a \emph{graded antihomomorphism}
on the algebra of modes. To emphasize the gradation we explicitly kept
the sign stemming from the anticommutation of the modes. Note that
there is no problem in commuting the modes since after acting on the
vertex they belong to different Hilbert spaces, so the only effect is
an additional sign. Finally, it is straightforward to check that
\begin{equation}
  \stl{12}\<V_2|V_2\>\stl{23} = \big( \uv\stl{3}\,\stl{1}\dvb+\dv\stl{3}
  \,\stl{1}\uvb\big) \exp\Big( \sfrac{1}{2}\sum_{m=1}^{\infty}\big[
  \ps^{+(3)}_{-m} \ps^{-(1)}_m + \ps^{-(3)}_{-m}\ps^{+(1)}_m\big] \Big) =
  \stl{3}\mathbbm{1}\stl{1} 
\end{equation}
by using standard coherent state techniques
(cf.~\cite{Kishimoto:2001ac,Bonora:2003xp,Kostelecky:2000hz})  and
eq.~(\ref{eq:V2bra}). One can then check that $\bpz\circ\bpz^{-1}=
\bpz^{-1}\circ\bpz=\mathbbm{1}$. This completes the construction of the
reflector state from the overlap equations.

\section{Interaction vertex} \label{sec:intvert}
\noindent
In this section, we set up the Neumann function method~\cite{Mandelstam:jk,
Mandelstam:1974fb, Mandelstam:1974fq, Kaku:xu, Cremmer:1974jq,
Cremmer:1974ej, LeClair:1988sp, LeClair:1988sj} for general $N$-string
vertices, since even in terms of the $\mathbbm{Z}_N$-%
Fourier-transforms the overlap equations are not directly soluble for
$N\geq 3$. In the case of the 3-string vertex, the Neumann coefficients
are computed explicitly in terms of generating functions. The observation
that they are intimately related to the well-known bosonic Neumann
coefficients helps us to show that the $K_n$-anomaly of the (bosonic and
fermionic) 3-vertex vanishes in any even dimension~$D$. Furthermore, it
will be shown that the 3-vertex for the $\ps^+\ps^-$ system satisfies its
overlap equations.

\noindent
{\bf Neumann function method.} The Neumann function method is based on
the fact that the large time transition amplitude is given by the Neumann
function of the scattering geometry under consideration. To find the Fock
space representation of the interaction vertex one makes an ansatz
quadratic in the oscillators,
\begin{equation}
  \<V_N| = {\cal{N}}_N \<\Om_N|\exp\big[ \sfrac{1}{4}\sum_{r,s}\sum_{k,l}
    \ps^{+(r)}_k N_{kl}^{rs} \ps^{-(s)}_l\big] \, , \label{eq:V_N}
\end{equation}
where ${\cal{N}}_N$ is a normalization factor which is determined
below.\footnote{The factor of of $\sfrac{1}{4}$ in the exponent will take
care of the nonstandard normalization of the correlation function for the
complex fermions.} The sum over the string labels $r$ and $s$ runs from 1
to $N$ and the restrictions on the summation range of the oscillator modes
has to be determined from the choice of vacuum $\<\Om_N|$ (cf.~(\ref{eq:Nvac}))
so that only creation operators appear in the vertex. As derived in
section~\ref{sec:refl}, $\ps^+_0$ does not occur in the exponential. 
The normalization factor ${\cal{N}}_N$ is determined by taking the
matrix element $\<V_N|\tilde\Om_N\>$ where $|\tilde\Om_N\>$ is the dual
vacuum $\<\Om_N|\tilde\Om_N\> = 1$. Since this matrix element corresponds
to a $\psi^+$ one-point function and $\psi^+$ has conformal weight zero
this yields ${\cal{N}}_N=\<V_N|\tilde\Om_N\>=\<\psi^+\>=1$.

To obtain an explicit expression for the coefficients we look at matrix
elements of the form
\begin{equation}
  G(z,w) = \<V_N|\psi^{+(s)}(z)\psi^{-(r)}(w)|\tilde\Om_N\> \label{eq:Green1}
\end{equation}
and reinterpret the result as a correlation function on the disk (or, thanks to
$PSL(2,\R)$ invariance, equivalently on the upper half-plane). Note that, in 
this expression, the $J$-anomaly has 
to be saturated in each string separately, i.\,e., in each Hilbert space 
we need one $\ps^+_0$ (which can be attributed to either $\<\Om_N|$ or
$|\tilde\Om_N\>$). 
Inserting the mode expansions for $\psi^+(z)$ and $\psi^-(w)$
into eq.~(\ref{eq:Green1}), one obtains by virtue of eq.~(\ref{eq:V_N})
\begin{equation}
  G(z,w)=\sum_{mn}z^n w^{m-1}N_{mn}^{rs}\, .
\end{equation}
Following~\cite{LeClair:1988sp}, we equate this with
\begin{equation}\label{eq:corr}
  G(z,w) = \< f_s\circ\ps^{+}(z) f_r\circ\ps^{-}(w)\sfrac{1}{N}
    \sum_{i=1}^N f_i\circ\psi^{+}(0)\>\,,
\end{equation}
where the sum on the right hand side was chosen to distribute the
background charge symmetrically among the $N$ strings. In principle,
any other choice of $\ps^+(0)$-insertions is admissible as long as
the $J$-anomaly on the scattering geometry is saturated, i.\,e., we
need a total $U(1)$ charge of $+1$ in the correlation function. 
The $f_r$ map the unit upper half-disk into the corresponding wedge of 
the scattering geometry, as defined in~(\ref{eq:fr}). The pole structure 
of the correlation function (\ref{eq:corr}) is easily evaluated; first order 
poles arise from $\ps^+\ps^-$-contractions, first order zeros from 
$\ps^+\ps^+$-contractions. Since the conformal weights of $\ps^+$ and 
$\ps^-$ are~0 and~1, respectively, we obtain
\begin{equation}
  \<f_s\circ\psi^{+}(z) f_r\circ\psi^{-}(w)\sfrac{1}{N}\sum_{i=1}^N
    f_i\circ\psi^{+}(0)\> = \frac{2f'_r(w)}{f_s(z)-f_r(w)}
    \frac{1}{N}\sum_{i=1}^N \frac{f_s(z)-f_i(0)}{f_r(w)-f_i(0)} \, .
    \label{eq:Green2}
\end{equation}
Here the unusual factor of $2$ appears due to the normalization of the
fermionic correlator. From eqs.~(\ref{eq:Green1}) to~(\ref{eq:Green2}) one
readily finds the expression for the Neumann coefficients in terms
of contour integrals,
\begin{equation}
\label{eq:Neumann}
  N_{mn}^{rs}=\oint\frac{dz}{2\pi\ic}\oint\frac{dw}{2\pi\ic}
    \frac{1}{z^{n+1}w^m}\frac{2f'_r(w)}{f_s(z)-f_r(w)}
    \frac{1}{N}\sum_{i=1}^N \frac{f_s(z)-f_i(0)}{f_r(w)-f_i(0)}\, .
\end{equation}

\noindent
{\bf Neumann coefficients and generating functions.} In this paragraph we
work out explicitly the integral formula for the Neumann coefficients for
the (bra-)interaction vertex and find expressions in terms of the coefficients
of generating functions. The vertex will take the form\footnote{%
Recall that no $\ps^{+(r)}_0$ appears in the exponent as substantiated in
section~\ref{sec:refl}.}
\begin{equation}
  \<V_3| = \big( \uudvb+\uduvb+\duuvb \big) \exp\big[ \sfrac{1}{4}\sum_{r,s}
    \sum_{k=1,l=0}^\infty \ps^{+(r)}_k N_{kl}^{rs} \ps^{-(s)}_l\big] \, .
    \label{eq:V3}
\end{equation}
The maps involved in~(\ref{eq:Neumann}) for $N=3$ can be gleaned
from~(\ref{eq:fr}),
\begin{equation}
  f_i(z)=e^{\sfrac{2\pi\ic}{3}(2-i)}\left(\frac{1+\ic z}{1-\ic z}
    \right)^{\sfrac{2}{3}}=\om^{2-i}f(z)
\end{equation}
with $\om=e^\frac{2\pi\ic}{3}$. Using these maps one can rewrite 
eq.~(\ref{eq:Neumann}) as
\begin{equation}
\label{eq:NCci}
\begin{split}
  N_{mn}^{rs} & = 2\oint\frac{dz}{2\pi\ic}\oint\frac{dw}{2\pi\ic}
    \frac{1}{z^{n+1}w^m}\frac{f_r'(w)}{f_s(z)-f_r(w)}
    \frac{f_s(z)f_r(w)^2-1}{f_r(w)^3-1} \\
  & = \oint\frac{dz}{2\pi\ic}\oint\frac{dw}{2\pi\ic}\frac{1}
    {z^{n+1}w^{m+1}}\frac{2}{3}\left(\frac{1}{1+zw}-\frac{w}{w-z}
    \right) \Big[ 1+U^{rs}(z,w)+U^{sr}(-z,-w)\Big] \, ,
\end{split}
\end{equation}
where
\begin{equation}
  U^{rs}(z,w) = \om^{(s-r)} \frac{w}{z}\left(\frac{1+\ic z}{1+\ic w}
    \right)^2 f(w)f(-z)\, .
\end{equation}
Introducing the generating functions
\begin{subequations}\label{eq:genfun}
\begin{align}
\label{eq:genfun1}
  G(z) & = \frac{f(z)}{(1+\ic z)^2} = \sum_{n=0}^{\infty}G_n z^n \, , \\
  H(z) & = (1+\ic z)^2f(-z) = \sum_{n=0}^{\infty}H_n z^n \, ,\label{eq:genfun2}
\end{align}
\end{subequations}
we can write
\begin{equation}\label{eq:NCUUbar}
 N_{mn}^{rs} = \frac{2}{3}\oint\frac{dz}{2\pi\ic}\oint\frac{dw}{2\pi\ic}
 \Big[C_{mn}(z,w)+\om^{(s-r)}U_{mn}(z,w)+\bar\om^{(s-r)}\bar U_{mn}(z,w)\Big]\,,
\end{equation}
with
\begin{subequations}
\begin{align}
  C_{mn}(z,w) & = \frac{1}{z^{n+1}w^{m+1}}\left(
    \frac{1}{1+zw}-\frac{w}{w-z}\right) \, , \\
  U_{mn}(z,w) & = \frac{1}{z^{n+2}w^{m}}\left(
    \frac{1}{1+zw}-\frac{w}{w-z}\right)G(w)H(z) \, , \\
  \bar U_{mn}(z,w) & = \frac{1}{z^{n+2}w^{m}}\left(
    \frac{1}{1+zw}-\frac{w}{w-z}\right)G(-w)H(-z)\, .
\end{align}
\end{subequations}
Performing the contour integrals\footnote{Note that one can choose the
contour always so that only the poles at zero contribute.}, one finds
the Neumann coefficients in terms of the coefficients of the generating
functions,
\begin{equation}
  N^{rs}_{mn} = \frac{2}{3} \left( C_{mn} + \om^{(s-r)} U_{mn} +
    \bar\om^{(s-r)} \bar U_{mn}\right)\, , \label{eq:NC}
\end{equation}
where
\begin{subequations}
\label{eq:CUUbar}
\begin{align}
  U_{mn} & = \sum_{k=0}^n \Big[ (-1)^{n+1-k} G_{m-n-2+k}-G_{m+n-k}
    \Big] H_k \, , \\
  \bar U_{mn} & = (-1)^{m+n}\sum_{k=0}^n \Big[ (-1)^{n+1-k}
    G_{m-n-2+k}-G_{m+n-k}\Big] H_k \, .
\end{align}
\end{subequations}
In these formulas, it is implicitly understood that coefficients with
negative index are zero, and, as usual, $C_{mn} = (-1)^m\de_{mn}$ for
$m,n>0$. Since $\ps^+$ does not appear in the exponential of the vertex,
we require $N^{rs}_{0m}=0$ for $m\geq 0$. Note that the Neumann
coefficients are real since the $G_n$ and $H_n$ are real for $n$ even
and purely imaginary for $n$ odd. Obviously, $\bar U$ is the complex
conjugate of $U$, and $\bar U = C U C$. Eq.~(\ref{eq:NC}) makes the
cyclic symmetry of the vertex manifest.

\noindent
{\bf Recursion relations.} To find recursion relations for the generating
functions~(\ref{eq:genfun}), we observe that $G(z)$ can be
expressed in terms of its derivative:
\begin{equation}
  G(z) = -\frac{3}{2}\frac{z^2+1}{3z+\ic} G'(z) \, .
\end{equation}
Inserting the mode expansion, one finds 
\begin{equation}
  G_k = -\oint\frac{dz}{2\pi\ic}\, \frac{1}{z^{k+1}}\frac{3}{2}
    \frac{z^2+1}{3z+\ic}\frac{\pa}{\pa z} G(z)\,.
\end{equation}
Partially integrating and evaluating the resulting contour integral leads 
to the following recursion formula for $G_n$:
\begin{equation}
  G_{k+2} = -\frac{2\ic}{3(k+2)}\, G_{k+1} - G_k \, . \label{eq:Grec}
\end{equation}
Note that this complies with the observation that the
$G_k$ are alternately real and imaginary. From~(\ref{eq:Grec})
and the initial condition $G_0 = G(0) = 1$ (and $G_{-1} := 0$), the
first coefficients are easily computed to be $G_1 = -\frac{2\ic}{3}$,
$G_2 = -\frac{11}{9}$, and $G_3 = \frac{76\ic}{81}$.

Similarly, we can use
\begin{equation}
  H(z) = \left( -\frac{\ic}{6} + \frac{z}{2} + \frac{4/3}{3z+\ic}
    \right) H'(z) \, 
\end{equation}
to find recursion relations for the $H_k$,
\begin{equation}
  (k+2) H_{k+2} = \frac{2\ic}{3} H_{k+1} - (k-2) H_k \, , \label{eq:Hrec}
\end{equation}
and with the initial condition $H_0 = H(0) = 1$ (and $H_{-1} := 0$),
the first coefficients are found to be $H_1 = \frac{2\ic}{3}$, $H_2
= \frac{7}{9}$ and $H_3 = \frac{32\ic}{81}$. One readily verifies
that
\begin{equation}
  \sum_{k=0}^n G_k H_{n-k} = 0 \qquad\text{for all }n\in\mathbbm{N} \, ,
\end{equation}
since $G(z) = 1/H(z)$.

\noindent
{\bf Relation to bosonic coefficients.} Exemplarily, the first few
Neumann coefficients $N^{11}_{mn}$ can be computed via eqs.~(\ref{eq:NC}),
(\ref{eq:CUUbar}) and the recursion relations~(\ref{eq:Grec})
and~(\ref{eq:Hrec}):
\begin{equation}
(N^{11})_{mn} = \begin{pmatrix}
  \sfrac{10}{27} & 0 & -\sfrac{64}{729} & 0 & \sfrac{832}{19683}
    & \ldots \\[1mm]
  0 & -\sfrac{26}{243} & 0 & \sfrac{1024}{19683} & 0 & \ldots \\[1mm]
  -\sfrac{64}{243} & 0 & \sfrac{1786}{19683} & 0 & -\sfrac{3008}{59049}
    & \ldots \\[1mm]
  0 & \sfrac{2048}{19683} & 0 & -\sfrac{10250}{177147} & 0 &
    \ldots \\[1mm]
  \sfrac{4160}{19683} & 0 & -\sfrac{15040}{177147} & 0 & \sfrac{82330}
    {1594323} & \ldots \\[1mm]
  \vdots & \vdots & \vdots & \vdots & \vdots & \ddots
\end{pmatrix}_{mn}\, .
\end{equation}
The above expression holds for $m, n\geq 1$. This suggests that, for
these values of $m, n$, the Neumann coefficients for the $\ps^+\ps^-$-%
system agree with those for the bosons in the momentum basis (cf.~eq.~%
(\ref{eq:bosN})) up to some factor; the same can be checked for all
other $r, s$:
\begin{equation}
  N^{rs}_{mn} = 2\sqrt{\frac{m}{n}} V^{rs}_{mn} \, . \label{eq:bosfer}
\end{equation}
A posteriori, one can easily find a proof for this relation.
Comparing with~(\ref{eq:bosN}) and~(\ref{eq:NCci}), we have to show that
\begin{equation}
  2\oint\frac{dz}{2\pi\ic}\oint\frac{dw}{2\pi\ic}\frac{1}{z^{n+1}
    w^m}\frac{f_r'(w)}{f_s(z)-f_r(w)}\frac{f_s(z)f_r(w)^2-1}
    {f_r(w)^3-1} = -\frac{2}{n}\oint\frac{dz}{2\pi\ic}\oint\frac{dw}
    {2\pi\ic}\frac{1}{z^n w^m}\frac{f_r'(w) f_s'(z)}{(f_s(z)-f_r(w))^2}\, .
    \label{eq:bosfer2}
\end{equation}
Since the right hand side can be rewritten as
\begin{equation}
  \frac{2}{n}\oint\frac{dz}{2\pi\ic}\oint\frac{dw}{2\pi\ic}\frac{1}
    {z^n w^m}\frac{\pa}{\pa z}\frac{f_r'(w)}{(f_s(z)-f_r(w))} =
    2\oint\frac{dz}{2\pi\ic}\oint\frac{dw}{2\pi\ic}\frac{1}
    {z^{n+1} w^m}\frac{f_r'(w)}{f_s(z)-f_r(w)} \, ,
\end{equation}
the difference of the left hand and the right hand sides of eq.~%
(\ref{eq:bosfer2}) is proportional to
\begin{equation}
  \oint\frac{dz}{2\pi\ic}\oint\frac{dw}{2\pi\ic}\frac{1}{z^{n+1} w^m}
    \frac{f_r'(w) f_r(w)^2}{f_r(w)^3-1}\, .
\end{equation}
This expression vanishes for $n>0$ due to the absence of poles in the
$z$-contour. This establishes the proof of eq.~(\ref{eq:bosfer}). \\

\noindent
{\bf Properties of the Neumann matrices.} In view of the close relation 
between the fermionic and 
the bosonic Neumann matrices one immediately obtains identities for
$N^{rs}_{mn}\,,\,m,n\ge 1$ from the bosonic ones. Defining $CN^{r\,r}=:N\,$, 
$\,CN^{r\,r+1}=:N_+$ and $CN^{r\,r-1}=:N_-$, one finds that 
$N$, $N^+$ and $N^-$ mutually commute and
\begin{gather}
  N+ N_+ + N_- = 2\, ,
  \quad N_+\,N_-=N(N-2)\, ,
  \quad N^2 + N_+^2 + N^2_- = 4\, , \notag \\
  N\,N_+ + N_+\,N_- + N_-\,N=0\, ,\qquad N^2_{\pm}-N_{\pm}= N\,N_{\mp}\, , \\
  C\,N = N\,C\,,\qquad C\,N_+=N_-\,C\, . \notag
\end{gather}
The proof of eq.~(\ref{eq:bosfer}) breaks down for $n=0$. We have 
$N^{rs}_{00}=0$; the Neumann coefficients for the case $n=0$ and $m>0$ 
are given by
\begin{align}
  N^{r\,r}_{m0} & = \begin{cases}
               -\frac{8\ic}{9m} G_{m-1} & \text{for $m$ even,} \\
               \quad 0 & \text{for $m$ odd,}
                  \end{cases} \\
  N^{r\,r+1}_{m0} & = \begin{cases}
               \frac{4\ic}{9m} G_{m-1} & \text{for $m$ even,} \\
               -\frac{4}{3\sqrt{3}m} G_{m-1} & \text{for $m$ odd,}
                  \end{cases} \\
  N^{r\,r-1}_{m0} & = \begin{cases}
               \frac{4\ic}{9m} G_{m-1} & \text{for $m$ even,} \\
               \frac{4}{3\sqrt{3}m} G_{m-1} & \text{for $m$ odd.}
                  \end{cases}
\end{align}
The indices $r,s$ are cyclic. From this it is obvious that
\begin{gather}
  C_{nm}N^{rt}_{m0}=N^{tr}_{n0}\,,\quad
  \sum_t\sum_m N^{rt}_{m0}=\sum_t\sum_mN^{tr}_{m0}=0\,.
\end{gather}
Exploiting the fact that the generating function $G(z)$ defined in 
eq.~(\ref{eq:genfun1}) is proportional to the
derivative of ${((1-\ic z)/(1+\ic z))^{1/3}}$, one can show that 
the coefficients $G_k$ are related to the coefficients $a_n$ (or 
equivalently $A_n$) defined in appendix~\ref{sec:bosN}~\cite{Gross:1986ia} via
\begin{equation}
  G_{m-1}=\sfrac{3}{2}m(-\ic)^{(m-1)}a_m\,.
\end{equation}
Evaluating the generating function for the coefficients $A_{2n}$ 
(cf. eq.~(\ref{eq:A_nB_n})),
\begin{equation}
  \frac{1}{2}\bigg[\Big(\frac{1-\ic\,z}{1+\ic\,z}\Big)^{1/3}
  +\Big(\frac{1+\ic\,z}{1-\ic\,z}\Big)^{1/3}\bigg]\,,
\end{equation}
at $z=1$, we obtain 
\begin{equation}
  \sum_{m=1}^{\infty}N^{11}_{m0}=\frac{4}{3}\sum_{n=1}^{\infty}A_{2n}=
  \frac{4}{3}\Big(\frac{\sqrt{3}}{2}-1\Big)\,.
\end{equation}
The contour integral around $z=0$ computes~\cite{Okuyama:2002yr}
\begin{equation}
  \sum_{n=0}^{\infty}(A_{2n}^2-A_{2n+1}^2)=
  \oint\frac{dz}{2\pi\ic}\frac{1}{z}
  \bigg(\frac{1+\ic\,z}{1-\ic\,z}\bigg)^{1/3}
  \bigg(\frac{1+\ic\,\sfrac{1}{z}}{1-\ic\,\sfrac{1}{z}}\bigg)^{1/3}
  =\frac{1}{2}\, ,
\end{equation}
which establishes that
\begin{equation}
  \sum_t\sum_{m=1}^{\infty} N^{1t}_{m0}N^{t1}_{m0}=
  \frac{8}{3}\Big(\sum_{n=0}^{\infty}(A_{2n}^2-A_{2n+1}^2)-1\Big)=
  -\frac{4}{3}\,.
\end{equation}

Having at hand fermion Neumann coefficients for the nonzero-modes expressed
in terms the boson Neumann coefficients puts us in the position
to compute the $K_n$-anomaly of the fermionic 3-vertex in a very simple
way. Similarly, the overlap equations can be checked more easily than
with the original expression~(\ref{eq:NC}). This will be done in the
next two paragraphs.

\noindent
{\bf Anomaly of the $\boldsymbol{\ps^\pm}$-vertex.} We will now demonstrate 
that the contribution of one $\ps^+\ps^-$-pair to the $K_n$-anomaly of the
3-vertex cancels the contribution of two real (or one complex) bosons. This
agrees with the fact that a $(1,0)$-first order system contributes $c=-2$ to 
the central charge. Thus, in contrast to bosonic and N=1 strings, no
restriction on the critical dimension follows from the $K_n$-anomaly. 

Namely, let $\sum_{r=1}^3K_m^{(r)\ps}=\sum_{r=1}^3
\big( L_{m}^{\ps(r)}- (-1)^mL_{-m}^{\ps(r)}\big)$ act on the 3-vertex~%
(\ref{eq:V3}). The only contribution to the $c$-number anomaly comes
from the terms in
\begin{equation}
 -(-1)^m L_{-m}^{\ps(r)} = -(-1)^m \frac{1}{2}\sum_k (m-k)\ps_{k-m}^{+(r)}\cdot
    \ps_{-k}^{-(r)}\,
\end{equation}
containing two creation operators, i.\,e., from $\frac{1}{2}
\sum_{k=0}^{m-1} (m-k)\ps_{k-m}^{+(r)}\cdot\ps_{-k}^{-(r)}$. The action
of $\ps_{k-m}^{+(r)}$ on the bra-vertex pulls down a sum over annihilation
operators, and from the interchange of $\ps_{-k}^{-(r)}$ with these creation
operators we get a $c$-number term:
\begin{equation}
\begin{split}
  \<V_3|K_m^{(3)\ps} & = -(-1)^m\frac{1}{2} \sum_{r=1}^3 \sum_{k=0}^{m-1}(m-k)
    \<V_3|\ps_{k-m}^{+(r)}\cdot\ps_{-k}^{-(r)} + \ldots \\
    & = -(-1)^m\frac{1}{4} \sum_{r,s=1}^3 \sum_{k=0}^{m-1}(m-k) 
       \sum_{l=0}^\infty
    \<V_3|N^{sr}_{l,m-k}\ps_l^{+(s)}\cdot\ps_{-k}^{-(r)} + \ldots \\
    & = -(-1)^m\frac{3\,D}{4}
      \sum_{k=1}^{m-1}(m-k) N^{rr}_{k,m-k}\<V_3|\, .
\end{split}
\end{equation}
In the third equality we have used that $N^{11}=N^{22}=N^{33}$ due to
cyclicity and that $N^{rr}_{0,m-k}=0$, i.e., the exponential in the
vertex contains no $\ps^+_0$. The dots indicate terms which do not
contribute to the $c$-number anomaly. From~(\ref{eq:bosanom}), this
equals twice the negative contribution of one real boson; the total
anomaly vanishes if we pair each $\ps^{+a}\ps^{-\ab}$-system with a
complex boson field $Z^a,\Zb^\ab$ in any even dimension.

\noindent
{\bf Overlap conditions.} According to the general method outlined in 
section \ref{sec:refl} we introduce $\Z_3$-Fourier-transforms 
\begin{subequations}\label{eq:Z3}
\begin{align}
\Psi^a = \frac{1}{\sqrt{3}}\sum_{r=1}^{3}\psi^{+(r)}\om^{ra}\,,
\\
\Pi^a = \frac{1}{\sqrt{3}}\sum_{r=1}^{3}\psi^{-(r)}\om^{ra}\,,
\end{align}
\end{subequations}
where $\om=e^\sfrac{2\pi\ic}{3}$ and the index $a$ runs from $1$ to $3$. 
This diagonalizes the overlap equations which then read
\begin{subequations}\label{eq:Z3ovlap1} 
\begin{align}
\<V_3|\Psi^1(\s)=\begin{cases}\om\,\<V_3|\Psi^1(\pi-\s)\,,
                              \quad\s\in[0\,,\sfrac{\pi}{2}]\,, \\
                              \bar\om\,\<V_3|\Psi^1(\pi-\s)\,,
                              \quad\s\in[\sfrac{\pi}{2},\pi]\,,
                 \end{cases} 
\\
\<V_3|\Psi^2(\s)=\begin{cases}\bar\om\,\<V_3|\Psi^2(\pi-\s)\,,
                              \quad\s\in[0\,,\sfrac{\pi}{2}]\,, \\
                              \om\,\<V_3|\Psi^2(\pi-\s)\,,
                              \quad\s\in[\sfrac{\pi}{2},\pi]\,,
                 \end{cases} 
\\
\<V_3|\Psi^3(\s)=\<V_3|\Psi^3(\pi-\s)\,,
\phantom{\quad\s\in[\sfrac{\pi}{2},\pi]}
\end{align}
\end{subequations}
and
\begin{subequations}\label{eq:Z3ovlap2} 
\begin{align}
\<V_3|\Pi^1(\s)=\begin{cases}-\om\,\<V_3|\Pi^1(\pi-\s)\,,
                              \quad\s\in[0\,,\sfrac{\pi}{2}]\,, \\
                             -\bar\om\,\<V_3|\Pi^1(\pi-\s)\,,
                              \quad\s\in[\sfrac{\pi}{2},\pi]\,,
                 \end{cases} 
\\
\<V_3|\Pi^2(\s)=\begin{cases}-\bar\om\,\<V_3|\Pi^2(\pi-\s)\,,
                              \quad\s\in[0\,,\sfrac{\pi}{2}]\,, \\
                             -\om\,\<V_3|\Pi^2(\pi-\s)\,,
                              \quad\s\in[\sfrac{\pi}{2},\pi]\,,
                 \end{cases} 
\\
\<V_3|\Pi^3(\s)=-\<V_3|\Pi^3(\pi-\s)\,.\phantom{\quad\s\in[\sfrac{\pi}{2},\pi]}
\end{align}
\end{subequations}
These overlap equations can be written in terms of the Fourier modes of the 
operator~\cite{Gross:1986ia}
\begin{equation}\label{eq:Y}
Y(\s,\s')=(-\sfrac{1}{2}+\sfrac{\sqrt{3}}{2}
[\ic\,\Theta(\sfrac{\pi}{2}-\s)-\ic\,\Theta(\s-\sfrac{\pi}{2})])
\de(\s+\s'-\pi)=:-\sfrac{1}{2}C(\s,\s')+\sfrac{\sqrt{3}}{2}X(\s,\s')
\end{equation}
as
\begin{subequations}\label{eq:Z3ovmod} 
\begin{align}
\sum_{l=0}^{\infty}(\tilde E_{kl}+\sfrac{1}{2}\tilde C_{kl}
-\sfrac{\sqrt{3}}{2}\tilde X_{kl})\<V_3|\tilde\Psi^1_l=0\,,
\\
\sum_{l=0}^{\infty}(\tilde E_{kl}+\sfrac{1}{2}\tilde C_{kl}
+\sfrac{\sqrt{3}}{2}\tilde X_{kl})\<V_3|\tilde\Psi^2_l=0\,,
\\
\sum_{l=0}^{\infty}(\tilde E_{kl}-\sfrac{1}{2}\tilde C_{kl}
+\sfrac{\sqrt{3}}{2}\tilde X_{kl})\<V_3|\tilde\Pi^1_l=0\,,
\\
\sum_{l=0}^{\infty}(\tilde E_{kl}-\sfrac{1}{2}\tilde C_{kl}
-\sfrac{\sqrt{3}}{2}\tilde X_{kl})\<V_3|\tilde\Pi^2_l=0\,,
\end{align}
\end{subequations}
where here and in the following the indices $k,l,j\in\mathbbm{N}_0$ while
$m,n\in\mathbbm{N}$. The matrices $\tilde E$ and $\tilde C$ are given by
\begin{equation}
\tilde E_{kl}=2\,\de_{0k}\de_{0l}+\de_{kl}\,,\quad
\tilde C_{kl}=(-1)^k\tilde E_{kl}\,.
\end{equation}
The matrices $\tilde X_{kl}$ can be found in appendix~\ref{sec:moreov}. The 
redefined oscillators are
$\tilde\Psi^a_m=\Psi^a_m+\Psi^a_{-m}$ and
$\tilde\Pi^a_m=\Pi^a_m+\Pi^a_{-m}$ for the nonzero-modes and
$\tilde\Psi^a_0=\Psi^a_0$ and $\tilde\Pi^a_0=\Pi^a_0$ for the zero-modes,
respectively.

We make the following ansatz for the interaction vertex in terms of the
$\Z_3$-transformed oscillators (recall the range of the indices
defined above!)
\begin{equation}\label{eq:V3ansatz}
\<V_3|=\<\Om_3|\exp\Big[\sfrac{1}{2}\sum_{m,n}\Psi_{m}^3 C_{mn}\Pi_{n}^3
                 +\sum_{m,k}(\Psi_{m}^2\tilde U_{mk}\Pi_{k}^1
                 +\Psi_{m}^1\tilde{\bar U}_{mk}\Pi_{k}^2)\Big]\,.
\end{equation}
Note that the exponential does not contain any $\Psi_0$ modes.
The vacuum $\<\Om_3|$ is given by
\begin{equation}\label{eq:Z3vac}
\<\Om_3|=\big\<\Pi_0^3=0,\Psi_0^1=0,\Psi_0^2=0\big|\,,
\end{equation}
which in terms of one string Hilbert space vacua is expressed as\footnote{%
The overlap equations fix the vacuum up to an overall sign factor.}
\begin{equation}\label{eq:V3vac}
\<\Om_3|=\stl{1}\uvb\otimes\stl{2}\uvb\otimes\stl{3}\dvb+
         \stl{1}\dvb\otimes\stl{2}\uvb\otimes\stl{3}\uvb+
         \stl{1}\uvb\otimes\stl{2}\dvb\otimes\stl{3}\uvb\,.
\end{equation}
It is straightforward to see that (\ref{eq:V3ansatz}) satisfies the overlap
equations for the $\Psi_m^3$'s and the $\Pi_m^3$'s. Comparing the
$\Z_3$-transformed version of the interaction vertex with
eq.~(\ref{eq:V3}), one can identify
\begin{equation}
\tilde U_{ml} = U_{ml}\,,\quad  \tilde{\bar U}_{ml} =\bar U_{ml}.
\end{equation}
Inserting the ansatz (\ref{eq:V3ansatz}) in (\ref{eq:Z3ovmod}), one obtains 
the overlap equations for the oscillators $\Pi^1$, $\Pi^2$ and 
$\Psi^1$, $\Psi^2$ in matrix form
\begin{subequations}\label{eq:Z3ovmatrix}
\begin{align}
\sum_{l=0}^{\infty}(\tilde E_{kl}-\sfrac{1}{2}\tilde C_{kl}
+\sfrac{\sqrt{3}}{2}\tilde X_{kl})(\de_{lj}-U_{lj})=0\,,\label{eq:ovPi1}
\\
\sum_{l=0}^{\infty}(\tilde E_{kl}-\sfrac{1}{2}\tilde C_{kl}
-\sfrac{\sqrt{3}}{2}\tilde X_{kl})(\de_{lj}-\bar U_{lj})=0\,,\label{eq:ovPi2}
\\
\sum_{l=0}^{\infty}(\tilde E_{kl}+\sfrac{1}{2}\tilde C_{kl}
-\sfrac{\sqrt{3}}{2}\tilde X_{kl})(\de_{lm}+\bar U^T_{lm})=0\,,
\label{eq:ovPsi1}
\\
\sum_{l=0}^{\infty}(\tilde E_{kl}+\sfrac{1}{2}\tilde C_{kl}
+\sfrac{\sqrt{3}}{2}\tilde X_{kl})(\de_{lm}+ U^T_{lm})=0\,.\label{eq:ovPsi2}
\end{align}
\end{subequations}
Let us now exemplify that these overlap conditions are indeed fulfilled by
the matrices given in eq.~(\ref{eq:CUUbar}). In particular we consider the
parts of the overlap equations involving zero-modes.

Consider the $k=0$ overlap equation for $\Pi^1_0$, which is the zero-zero
component of eq.~(\ref{eq:ovPi1}):
\begin{equation}\label{eq:ov00}
1-\sfrac{\sqrt{3}}{2}\sum_{m=1}^{\infty}\tilde X_{0m}U_{m0}\stackrel{!}{=}0\,.
\end{equation}
Inserting the $U_{m0}$ component
\begin{equation}
U_{m0}=-(-\ic)^m\,a_m
\end{equation}
into (\ref{eq:ov00}) allows us to use known summation
formulas for the coefficients~\cite{Gross:1986ia,Ohta:wn,Bordes:1993cd}
to obtain
\begin{equation}
\sum_{m=1}^{\infty}\tilde X_{0m}U_{m0}
=\frac{4}{\pi}\sum_{k=0}^{\infty}\frac{a_{2k+1}}{2k+1}=\frac{2}{\sqrt{3}}
\end{equation}
proving eq.~(\ref{eq:ov00}). Consider now the overlap equations for $k\ne0$. 
Setting $k=2l$ for $k$ even and $k=2l+1$ for $k$ odd yields 
\begin{subequations}\label{eq:ovk0} 
\begin{align}
&-\sfrac{1}{2}U_{2l,0}
-\sfrac{\sqrt{3}}{2}\sum_{m=1}^{\infty}
          \tilde X_{2l,m}U_{m0}\stackrel{!}{=}0\,, \\
&-\sfrac{3}{2}U_{2l+1,0}+\sfrac{\sqrt{3}}{2}\tilde X_{2l+1,0}
-\sfrac{\sqrt{3}}{2}\sum_{m=1}^{\infty}
          \tilde X_{2l+1,m}U_{m0}\stackrel{!}{=}0\,.
\end{align}
\end{subequations}
The first of these equations is proven by 
\begin{equation}
\frac{\sqrt{3}}{2}\sum_{m=1}^{\infty}\tilde X_{2l,m}U_{m0}=
\frac{\sqrt{3}}{\pi}\sum_{k=0}^{\infty}(-1)^l
\left(\frac{a_{2k+1}}{2k+1+2l} + \frac{a_{2k+1}}{2k+1-2l}\right)
=\frac{1}{2}(-1)^l a_{2l}\,.
\end{equation}
The second equation in~(\ref{eq:ovk0}) is fulfilled due to 
\begin{align}
\frac{\sqrt{3}}{2}\sum_{m=1}^{\infty}\tilde X_{2l+1,m}U_{m0}&=
-\frac{\sqrt{3}\ic}{\pi}\sum_{k=0}^{\infty}(-1)^l
\left(\frac{a_{2k}}{2k+2l+1} - \frac{a_{2k}}{2k-2l-1}\right)
+\frac{\sqrt{3}\ic}{\pi}(-1)^l\frac{2\,a_0}{2k+1} \notag \\
\phantom{\frac{\sqrt{3}}{2}\sum_{m=1}^{\infty}\tilde X_{2l+1,m}U_{m0}}
&=-\frac{3\ic}{2}(-1)^l a_{2l+1}
+\frac{\sqrt{3}\ic}{\pi}(-1)^l\frac{2\,a_0}{2k+1}\,.
\end{align}
More involved overlap conditions can be proven using techniques developed in 
~\cite{Gross:1986ia,Ohta:wn,Bordes:1993cd}. We postpone their discussion to  
appendix \ref{sec:moreov}.  

\section{Conclusions}
\noindent
In this paper we explicitly constructed the string field theory vertices 
for a fermionic first order system $\psi^{\pm}$ with conformal weights 
$(1,0)$ in the operator formulation. The technical ingredients needed to 
construct general $N$-string vertices are presented in detail. 
The identity vertex, the reflector and the interaction vertex are discussed 
with emphasis on their charge under the anomalous $U(1)$ current $J$ and 
their zero-mode dependence. The identity vertex 
and the reflector are derived from the corresponding $\de$-function overlap 
conditions. The reflector is shown to implement BPZ conjugation as a graded 
antihomomorphism, and some consistency conditions on the gluing of the 
reflector are checked. The construction of the interaction vertex is achieved 
by invoking the Neumann function method. The coefficients of the Neumann 
matrices are given in terms of coefficients of generating functions and 
recursion relations for these coefficients are derived. 
The Neumann coefficients for the $\psi^{\pm}$ system are 
neatly expressed in terms of those for the bosons. 
This allows us to infer identities for the fermion 
Neumann matrices directly from those for the bosons. Moreover, the 
$c$-number anomaly of midpoint preserving reparametrizations for a 
$\psi^{\pm}$ pair is straightforwardly shown to cancel the contribution of 
two real bosons. This agrees with the fact that a $(0,1)$ first order system 
contributes $c=-2$ to the central charge. Eventually, it is shown that the 
overlap equations following from the $\de$-function overlap conditions are
satisfied by the Neumann matrices. 

Clearly, the work presented is meant to be a starting point for further 
investigations. Diagonalizing the vertex is a straightforward task and is 
attacked in a forthcoming paper~\cite{Ihl:200nie}. This should pave the way 
for studying solutions to string field theory in several contexts. Firstly,
one might examine how the solution generating techniques derived from 
integrability~\cite{Lechtenfeld:2000qj} and proposed
in~\cite{Lechtenfeld:2002cu,Kling:2002ht,Kling:2002vi} perform in the more
controlled setting of $N=2$ SFT. This can be expected to give valuable 
information about how solutions to string field theory can be constructed 
dropping the factorization assumption of vacuum string field theory. 
As a direct application to N=1 superstring field theory it appears to be 
worthwhile to investigate the dependence of solutions on the $\eta\xi$-system 
more closely. This fermionic $(0,1)$ first order system emerges in the 
bosonization of the superconformal ghosts. The related picture changing 
operation is a delicate subject in string (field) theory and deserves 
to be examined with minuteness. Finally, we want to point out the similarity 
of the fermionic first order system considered in this paper and twisted 
$bc$-system. This auxiliary boundary conformal field theory is used in 
vacuum string field theory to construct solutions to the ghost part of the 
equation of motion. The solutions to these equations become projectors in 
the {\it twisted} theory. It is tempting to speculate that this similarity can 
be traced to a deeper interrelation.

\subsubsection*{Acknowledgements} 
\noindent
We would like to thank O. Lechtenfeld, A. D. Popov, M.Ihl and M. Wolf for 
useful discussions. This work was done within the framework of the DFG 
priority program ``String Theory'' (SPP 1096).
\\ 

\begin{appendix}

\section{Review of bosonic results}\label{sec:bos}
\noindent
In this section we briefly review the results on bosonic vertices
in string field theory which will be useful in section~\ref{sec:intvert}. 
The operator formulation of Witten's open string field theory has been 
developed in~\cite{Gross:1986ia,Ohta:wn,Gross:1986fk,Samuel:1986wp} for the bosonic 
string and in~\cite{Gross:1987pp} for the NSR superstring. 

\noindent 
{\bf Bosonic vertices.} Using the open string mode expansion and corresponding 
creation and annihilation operators\footnote{Strictly speaking the formulas
given in the following are valid for one single boson but straightforwardly 
generalizable to any number of spacetime directions by introducing the 
corresponding Lorentz indices.}
\begin{equation}
X(\s)=x_0+\sqrt{2}\sum_{m=1}^{\infty}x_n\,\cos n\s\quad\text{with}
\quad x_n=\ic\sqrt{\sfrac{\a'}{n}}(a_n-a_n^{\dagger})\,,
\quad x_0=\ic\sqrt{\sfrac{\a'}{2}}(a_0-a_0^{\dagger})\,, 
\end{equation}
one obtains the identity vertex\footnote{Originally, in~\cite{Gross:1986ia} 
the corresponding ket-vector has been constructed. For coherence with the 
rest of the presentation we review the bosonic results in terms of 
bra-vectors.}
\begin{equation}
X(\s)=X(\pi-\s)\quad\Rightarrow\quad
\<V_1|=\<\Ical|=
\<0|\exp\Big[-\sfrac{1}{2}
\sum_{k,l\ge0}^{\infty}a_k C_{kl}\,a_k\Big]\,,
\end{equation}
where $C_{kl}=(-1)^k\de_{kl}$. 
The bra-vacuum used in this expression is the oscillator vacuum which is 
annihilated by all $a_k^{\dagger}$. The 2-string overlap conditions  
\begin{equation}
X^{(1)}(\s)=X^{(2)}(\pi-\s) \quad\Rightarrow\quad 
\<V_2|\,x_n^{(1)}=(-1)^n\,\<V_2|\,x_n^{(2)}\,,
\end{equation}
are solved in terms of the same matrix $C$ by
\begin{equation}
\<V_2|=
\<0|\exp\Big[-\sfrac{1}{2}
\sum_{k,l\ge0}^{\infty}a^{(1)}_kC_{kl}\,a^{(2)}_l\Big]\,.
\end{equation}
The construction of the interaction vertex $\<V_3|$ is more involved, and 
the overlap equations are not solved directly. 
They are most conveniently formulated in terms of $\Z_3$-transformed string 
oscillators $A_k^{(a)}$ (cf. section~\ref{sec:refl}, not to be confused 
with the coefficients $A_k$ defined in~(\ref{eq:A_nB_n})). 
By using 
\begin{equation}
\<V_3|=\<0|\exp\Big[-
\sum_{k,l\ge0}^{\infty}\Big(\sfrac{1}{2}A^{(3)}_kC_{kl}\,A^{(3)}_l
+A^{(1)}_kU_{kl}\,A^{(2)}_l\Big)\Big]
\end{equation}
as an ansatz for the vertex, one can write the overlap equations in matrix 
form as
\begin{equation}
(\mathbbm{1}-Y)E(\mathbbm{1}+U)=0\,,\quad 
(\mathbbm{1}+Y)E^{-1}(\mathbbm{1}-U)=0\,.
\end{equation}
Here the matrix $E$ has components 
$E_{mn}=\de_{m,0}\de_{n,0}+\sqrt{\sfrac{2}{n}}\de_{mn}\,$, 
and the matrix $Y$ is given by the Fourier components of the 
operator~(\ref{eq:Y}). Rewritten in terms of the original one-string 
oscillators the vertex takes the form\footnote{We stick to the 
notation of~\cite{Rastelli:2001jb} and denote the matrices in the 
oscillator basis with a prime while those in momentum basis are unprimed.}
\begin{equation}
\<V_3|=
\<0|\exp\Big[-\sfrac{1}{2}\sum_{r,s}
\sum_{m,n\ge0}^{\infty}a^{(r)}_n{V'}^{rs}_{nm}\,a^{(s)}_m\Big]\,
\end{equation}
with the matrices ($r=1,2,3$)
\begin{subequations}
\begin{align}
& {V'}^{r\,r}_{nm}=\sfrac{1}{3}(C+U+\bar U)\,, \\
& {V'}^{r\,r+1}_{nm}=
\sfrac{1}{3}(C-\sfrac{1}{2}(U+\bar U)+\sfrac{\sqrt{3}}{2}(U-\bar U))\,, \\
& {V'}^{r\,r-1}_{nm}= 
\sfrac{1}{3}(C-\sfrac{1}{2}(U+\bar U)-\sfrac{\sqrt{3}}{2}(U-\bar U))\,.
\end{align} 
\end{subequations}
After transformation to momentum basis, these matrices can be identified with 
the Neumann coefficients. 
The explicit coefficients are listed in appendix \ref{sec:bosN}. Following 
the conformal field theory approach of~\cite{LeClair:1988sp}, one can express 
the Neumann coefficients via contour integrals 
\begin{equation}\label{eq:bosN}
V_{mn}^{rs}=-\frac{1}{\sqrt{mn}}\oint\frac{dz}{2\pi\ic}\oint\frac{dw}{2\pi\ic}
\frac{1}{z^n\,w^m}\frac{f_r'(w)f_s'(z)}{(f_s(z)-f_r(w))^2}\,,
\end{equation}
where the $f_i$'s map the scattering geometry of the interaction vertex to 
the disk (cf. section~\ref{sec:id}). 

\noindent
{\bf Reparametrization anomaly.} Reparametrizations generated by 
$K_n=L_n-(-1)^nL_{-n}$ leave the midpoint invariant and are classical  
symmetries of string field theory. However, for nonvanishing central charge 
one finds an anomaly in these reparametrizations. In the operator formulation 
this anomaly arises from operator orderings when two creation operators 
act on the vertex, i.e., from terms like
\begin{equation}
-(-1)^m\sfrac{1}{2}
\sum_{k=1}^{m-1}\sqrt{k(m-k)}a_k^{\dagger}\cdot a_{m-k}^{\dagger}
\end{equation}
contained in $K_m$. In~\cite{Gross:1986fk} it was shown that a single boson 
contributes (see also~\cite{Schnabl:2002ff}) 
\begin{equation}
\<V_3|(K_{2n}^{(1)}+K_{2n}^{(2)}+K_{2n}^{(3)})=-\sfrac{5}{18}n(-1)^n\<V_3|\,.
\end{equation}
However, due to a nontrivial relation between the Neumann coefficients for 
the bosons and the fermions in the twisted theory we can show that in the 
full theory this anomaly is canceled in any (even) dimension. To this end let 
us derive the contribution of $D$ bosons to the anomaly in terms of the 
Neumann coefficients. The application of $\sum_{t=1}^3K_{m}^{(t)}$ to $\<V_3|$ 
yields
\begin{align}\label{eq:bosanom}
\begin{split}
\sum_{t=1}^3\<V_3|K_{m}^{(t)}
&=-(-1)^m\frac{1}{2}\sum_{t=1}^3\sum_{k=1}^{m-1}
   \sqrt{k(m-k)}\<V_3|a_k^{\dagger(t)}\cdot a_{m-k}^{\dagger(t)}+\dots \\
&=(-1)^m\frac{1}{4}\sum_{t=1}^3\sum_{k=1}^{m-1}\sqrt{k(m-k)}
    \<V_3|a_n^{(r)}(V_{nk}^{rt}+V_{kn}^{tr})\cdot a_{m-k}^{\dagger(t)}+\dots \\
&=(-1)^m\frac{D}{4}\sum_{t=1}^3\sum_{k=1}^{m-1}\sqrt{k(m-k)}
    \<V_3|(V_{m-k,k}^{tt}+V_{k,m-k}^{tt}) \\
&=(-1)^m\frac{3\,D}{2}
    \sum_{k=1}^{m-1}\sqrt{k(m-k)}\,V_{k,m-k}^{tt}\<V_3| \,.
\end{split}
\end{align}

\section{Bosonic Neumann coefficients}\label{sec:bosN}
\noindent
In this appendix we give a list of the boson Neumann coefficients. These are 
results of~\cite{Gross:1986ia,Ohta:wn} but presented in the notation 
of~\cite{Rastelli:2001jb}. The interaction vertex in momentum basis is given 
by 
\begin{equation}
\<V_3|=\int d^Dp^{(1)}d^Dp^{(2)}d^Dp^{(3)}\de^D(p^{(1)}+p^{(2)}+p^{(3)})
       \stl{123}\<p,0|\exp \big[-V\big]\,,
\end{equation}
where
\begin{equation}
V=\frac{1}{2}
\sum_{r,s}\sum_{m,n\ge 1}\h_{\mu\nu}a_m^{(r)\mu}V_{mn}^{rs}a_n^{(s)\nu} + 
\sap
\sum_{r,s}\sum_{n\ge 1}\h_{\mu\nu}p^{(r)\mu}V_{0n}^{rs}a_n^{(s)\nu}+
\frac{\a'}{2}\sum_{r}\h_{\mu\nu}p^{(r)\mu}V_{00}^{rr}p^{(r)\nu}\,.
\end{equation}
The Neumann matrices are expressed in terms of 
coefficients $A_n$ and $B_n$ which are defined as 
\begin{equation}\label{eq:A_nB_n}
\left(\frac{1+\ic\,z}{1-\ic\,z}\right)^{1/3}=
\sum_{n\text{ even}}A_n\,z^n+\ic\sum_{n\text{ odd}}A_n\,z^n\,,\qquad
\left(\frac{1+\ic\,z}{1-\ic\,z}\right)^{2/3}=
\sum_{n\text{ even}}B_n\,z^n+\ic\sum_{n\text{ odd}}B_n\,z^n\,.
\end{equation}
The Neumann coefficients read 
\begin{subequations}
\begin{align}
V_{mn}^{r\,r}&=-\sqrt{mn}\,\frac{(-1)^n+(-1)^{m}}{6}
    \Big(
    \frac{A_m\,B_n+A_n\,B_m}{m+n}
   +\frac{A_m\,B_n-A_n\,B_m}{m-n}\Big)\,,\quad m\ne n,~m,n\ne 0\,, 
\\
V_{mn}^{r\,r+1}&=\sqrt{mn}\,\frac{(-1)^n+(-1)^{m}}{12}\Big(
    \frac{A_m\,B_n+A_n\,B_m}{m+n}
   +\frac{A_m\,B_n-A_n\,B_m}{m-n}\Big) \notag\\
&-\sqrt{mn}\,\sqrt{3}\,\frac{1-(-1)^{n+m}}{12}\Big(
    \frac{A_m\,B_n-A_n\,B_m}{m+n}
   +\frac{A_m\,B_n+A_n\,B_m}{m-n}\Big)\,,\quad m\ne n,~m,n\ne 0\,, \\
V_{mn}^{r\,r-1}&=\sqrt{mn}\,\frac{(-1)^n+(-1)^{m}}{12}\Big(
    \frac{A_m\,B_n-A_n\,B_m}{m-n}
   +\frac{A_m\,B_n+A_n\,B_m}{m+n}\Big) \notag\\
&+\sqrt{mn}\,\sqrt{3}\,\frac{1-(-1)^{n+m}}{12}\Big(
    \frac{A_m\,B_n+A_n\,B_m}{m-n}
   +\frac{A_m\,B_n-A_n\,B_m}{m+n}\Big)\,,\quad m\ne n,~m,n\ne 0\,,
\end{align}
\end{subequations}
The coefficients on the diagonal are given by 
\begin{subequations}
\begin{align}
V^{r\,r}_{nn}&=-\frac{1}{3}\big[
         2\sum_{k=0}^n(-1)^{n-k}A_k^2-(-1)^n-A_n^2\big]\,,\quad n\ne 0\,,\\
V^{r\,r+1}_{nn}&=V^{r,r-1}_{nn}=
          \frac{1}{2}\big[(-1)^n-V^{rr}_{nn}\big]\,,\quad n\ne 0\,,\\
V^{r\,r}_{00}&=\ln (27/16)\,.
\end{align}
\end{subequations}
The coefficients with one index zero are obtained as limits of the 
coefficients $V_{mn}$ above
\begin{subequations}
\begin{align}
V^{r\,r}_{0n}&=-\sqrt{\frac{2}{n}}\,\frac{1+(-1)^n}{3}A_n\,, \\
V^{r\,r+1}_{0n}&=-\sqrt{\frac{2}{n}}\,\Big[-\frac{1+(-1)^n}{6}A_n
                 -\sqrt{3}\frac{1-(-1)^n}{6}A_n\Big]\,, \\
V^{r\,r-1}_{0n}&=-\sqrt{\frac{2}{n}}\,\Big[-\frac{1+(-1)^n}{6}A_n
                      +\sqrt{3}\frac{1-(-1)^n}{6}A_n\Big]\,.
\end{align}
\end{subequations}
The value of the coefficients for different conventions for $\a'$ are 
easily obtained absorbing the explicit $\a'$ into these coefficients. 

\section{More overlap equations}\label{sec:moreov}
\noindent
In this appendix we continue the discussion of the overlap equations started 
at the end of section~\ref{sec:intvert}. We adopt the convention about the 
index range chosen there so that the indices $k,l$ and $j$ start from zero, 
$k,l,j=0,1,\dots$, while $m,n=1,2,\dots\,$. The matrices $\tilde X_{kl}$ are
\begin{subequations}
\begin{align}
\tilde X_{0m}&=-\tilde X_{m0}=
\frac{2\,\ic}{\pi\,m}(-1)^{\sfrac{m-1}{2}}[1-(-1)^m]\,, \label{eq:X0}\\
\tilde X_{nm}&=\frac{\ic}{\pi}(-1)^{\sfrac{n-m-1}{2}}[1-(-1)^{n+m}]
\Big[\frac{1}{n+m}+\frac{(-1)^m}{n-m}\Big]\,. \label{eq:X}
\end{align} 
\end{subequations}
Note that compared to the matrices defined in~\cite{Gross:1986ia} 
we have $\tilde X_{nm}=X_{nm}^{\text{GJ}}$ but for the parts containing 
a zero index $-\sqrt{2}\,\tilde X_{0m}=X_{0m}^{\text{GJ}}$! Using the 
relation to the bosonic coefficients given in eq.~(\ref{eq:bosfer}) 
and the definition of $N^{r,r+1}_{mn}$ in~(\ref{eq:NC}) one finds for $m\ne n$
\begin{subequations}
\begin{align}
N^{r\,r}_{mn}&=\frac{2}{3}(U_{mn}+\bar U_{mn})
=2\sqrt{\frac{m}{n}} V^{r\,r}_{mn} \,,
\\
N^{r\,r+1}_{mn}&=\frac{2}{3}(\om U_{mn}+\bar\om\bar U_{mn})
=-\frac{1}{3}(U_{mn}+\bar U_{mn})+\frac{\ic}{\sqrt{3}}(U_{mn}-\bar U_{mn})
=2\sqrt{\frac{m}{n}} V^{r\,r+1}_{mn} \,,
\end{align} 
\end{subequations}
and hence
\begin{align}
\begin{split}
\label{eq:U}
U_{mn}=&-\frac{m}{4}\big[(-1)^n+(-1)^{m}\big]
    \Big[
    \frac{A_m\,B_n+A_n\,B_m}{m+n}
   +\frac{A_m\,B_n-A_n\,B_m}{m-n}\Big] \\
\phantom{U_{mn}=}&+\frac{\ic\,m}{4}\big[1-(-1)^{m+n}\big]
    \Big[\frac{A_m\,B_n-A_n\,B_m}{m+n}
   +\frac{A_m\,B_n+A_n\,B_m}{m-n}\Big]\,,\quad m\ne n\,,
\end{split}
\end{align}
from which it is once more apparent that $CUC=\bar U$. This prepares the 
stage to scrutinize the overlap equations for $\Pi^1_m$ following 
from eq.~(\ref{eq:ovPi1}). Taking $k=0$ and $j=2l$ in eq.~(\ref{eq:ovPi1}), 
one finds
\begin{equation}\label{eq:OV1}
\sum_{m=1}^{\infty}\tilde X_{0m}U_{m,2l}\stackrel{!}{=}0\,.
\end{equation}
Inserting eqs.~(\ref{eq:X0}) and~(\ref{eq:U}) yields
\begin{equation}
\sum_{m=1}^{\infty}\tilde X_{0m}U_{m,2l}=-\frac{2}{\pi}
\sum_{k=0}^{\infty}(-1)^k\Big[\frac{A_{2k+1}\,B_{2l}-A_{2l}\,B_{2k+1}}
{2k+1+2l}+\frac{A_{2k+1}\,B_{2l}+A_{2l}\,B_{2k+1}}{2k+1-2l}\Big]\,, 
\end{equation}
Using the relation between the coefficients $A_n$ and $a_n$,
\begin{equation}
A_{2k}=(-1)^k a_{2k}\,,\qquad A_{2k+1}=(-1)^k a_{2k+1}\,,
\end{equation}
and the summation formulas for the coefficients $a_n$ derived 
in~\cite{Gross:1986ia}
\begin{subequations}\label{eq:sumsums}
\begin{align}
O^a_k&=\sum_{l=0}^{\infty}\frac{a_{2l+1}}{(2l+1)+k}=
\frac{\pi\,a_k}{\sqrt{3}}\,,
&
O^a_{-n}&=\sum_{l=0}^{\infty}\frac{a_{2l+1}}{(2l+1)-n}=
-\frac{1}{2}\frac{\pi\,a_n}{\sqrt{3}}\,, &&\text{for $k,n$ even,}
\\
E^a_k&=\sum_{l=0}^{\infty}\frac{a_{2l}}{(2l)+k}=
\frac{\pi\,a_k}{\sqrt{3}}\,,
&
E^a_{-n}&=\sum_{l=0}^{\infty}\frac{a_{2l+1}}{(2l+1)-n}=
-\frac{1}{2}\frac{\pi\,a_n}{\sqrt{3}}\,, &&\text{for $k,n$ odd,}
\\
O^b_k&=\sum_{l=0}^{\infty}\frac{b_{2l+1}}{(2l+1)+k}=
\frac{\pi\,a_k}{\sqrt{3}}\,,
&
O^b_{-n}&=\sum_{l=0}^{\infty}\frac{b_{2l+1}}{(2l+1)-n}=
\frac{1}{2}\frac{\pi\,b_n}{\sqrt{3}}\,, &&\text{for $k,n$ even,}
\\
E^b_k&=\sum_{l=0}^{\infty}\frac{b_{2l}}{(2l)+k}=
\frac{\pi\,b_k}{\sqrt{3}}\,,
&
E^b_{-n}&=\sum_{l=0}^{\infty}\frac{b_{2l+1}}{(2l+1)-n}=
\frac{1}{2}\frac{\pi\,b_n}{\sqrt{3}}\,, &&\text{for $k,n$ odd,}
\end{align}
\end{subequations}
one finds
\begin{equation}
\sum_{m=1}^{\infty}\tilde X_{0m}U_{m,2l}=-\frac{2}{\pi}
\big[O^a_{2l}B_{2l}-O^b_{2l}A_{2l}+O^a_{-2l}B_{2l}+O^b_{-2l}A_{2l}\big]=0
\end{equation}
proving~(\ref{eq:OV1}). Now let us look at $k=2l+1$ and $j=2n+1$ 
in eq.~(\ref{eq:ovPi1}):
\begin{equation}\label{eq:OV2}
\sfrac{3}{2}U_{2l+1,2n+1}
+\sfrac{\sqrt{3}}{2}\sum_{m=0}^{\infty}\tilde X_{2l+1,m}U_{m,2n+1}
\stackrel{!}{=}0.
\end{equation}
By inserting eqs.~(\ref{eq:X}) and~(\ref{eq:U}), the sum can be written 
in terms of eqs.~(\ref{eq:sumsums}) as
\begin{align}
\begin{split}
\sum_{m=0}^{\infty}\tilde X_{2l+1,m}U_{m,2n+1}=&
(-1)^l\frac{2l+1}{\pi}\Big[
\frac{E^a_{2l+1}B_{2n+1}-E^b_{2l+1}A_{2n+1}}{(2n+1)-(2l+1)}+
\frac{E^a_{-2l-1}B_{2n+1}-E^b_{-2l-1}A_{2n+1}}{(2n+1)+(2l+1)}
\\
&
-\frac{E^b_{-2l-1}A_{2n+1}+E^a_{-2l-1}B_{2l+1}}{(2n+1)-(2l+1)}
-\frac{E^a_{2l+1}B_{2n+1}+E^b_{2l+1}A_{2n+1}}{(2n+1)+(2l+1)}\Big]
\\
=&-\sqrt{3}\,U_{2l+1,2n+1}\,,
\end{split}
\end{align}
proving eq.~(\ref{eq:OV2}). The case $k=2l$ and $j=2n$ can be easily treated 
along the same lines. 

\end{appendix}


\vspace*{5mm}

\begin{thebibliography}{99}
\bibitem{Witten:1985cc}
E.~Witten,
{\it Noncommutative geometry and string field theory},
Nucl.\ Phys.\ B {\bf 268} (1986) 253.

\bibitem{Sen:1999mh}
A.~Sen,
{\it Descent relations among bosonic D-branes}, 
Int.\ J.\ Mod.\ Phys.\ A {\bf 14} (1999) 4061
[arXiv:hep-th/9902105].

\bibitem{Sen:1999mg}
A.~Sen,
{\it Non-BPS states and branes in string theory},
arXiv:hep-th/9904207.

\bibitem{Sen:1999xm}
A.~Sen,
{\it Universality of the tachyon potential}, 
JHEP {\bf 9912} (1999) 027
[arXiv:hep-th/9911116].

\bibitem{Ohmori:2001am}
K.~Ohmori,
{\it A review on tachyon condensation in open string field theories},
arXiv:hep-th/0102085.

\bibitem{DeSmet:2001af}
P.~J.~De Smet,
{\it Tachyon condensation: Calculations in string field theory},
arXiv:hep-th/0109182.

\bibitem{Arefeva:2001ps}
I.~Ya.~Aref'eva, D.~M.~Belov, A.~A.~Giryavets, A.~S.~Koshelev 
and P.~B.~Medvedev,
{\it Noncommutative field theories and (super)string field theories},
arXiv:hep-th/0111208.

\bibitem{Taylor:2002uv}
W.~Taylor,
{\it Lectures on D-branes, tachyon condensation, and string field theory},
arXiv:hep-th/0301094.

\bibitem{Rastelli:2000hv}
L.~Rastelli, A.~Sen and B.~Zwiebach,
{\it String field theory around the tachyon vacuum},
Adv.\ Theor.\ Math.\ Phys.\  {\bf 5} (2002) 353
[arXiv:hep-th/0012251].

\bibitem{Rastelli:2001jb}
L.~Rastelli, A.~Sen and B.~Zwiebach,
{\it Classical solutions in string field theory around the tachyon vacuum},
Adv.\ Theor.\ Math.\ Phys.\  {\bf 5} (2002) 393
[arXiv:hep-th/0102112].

\bibitem{Rastelli:2001rj}
L.~Rastelli, A.~Sen and B.~Zwiebach,
{\it Half strings, projectors, and multiple D-branes in vacuum string field  theory},
JHEP {\bf 0111} (2001) 035
[arXiv:hep-th/0105058].

\bibitem{Gross:2001rk}
D.~J.~Gross and W.~Taylor,
{\it Split string field theory. I},
JHEP {\bf 0108} (2001) 009
[arXiv:hep-th/0105059].

\bibitem{Gross:2001yk}
D.~J.~Gross and W.~Taylor,
{\it Split string field theory. II},
JHEP {\bf 0108} (2001) 010
[arXiv:hep-th/0106036].

\bibitem{Schnabl:2002ff}
M.~Schnabl,
{\it Anomalous reparametrizations and butterfly states in string field theory},
Nucl.\ Phys.\ B {\bf 649} (2003) 101
[arXiv:hep-th/0202139].

\bibitem{Gaiotto:2002kf}
D.~Gaiotto, L.~Rastelli, A.~Sen and B.~Zwiebach,
{\it Star algebra projectors},
JHEP {\bf 0204} (2002) 060
[arXiv:hep-th/0202151].

\bibitem{Fuchs:2002zz}
E.~Fuchs, M.~Kroyter and A.~Marcus,
{\it Squeezed state projectors in string field theory},
JHEP {\bf 0209} (2002) 022
[arXiv:hep-th/0207001].

\bibitem{Chen:2002jd}
B.~Chen and F.~L.~Lin,
{\it D-branes as GMS solitons in vacuum string field theory},
Phys.\ Rev.\ D {\bf 66} (2002) 126001
[arXiv:hep-th/0204233].

\bibitem{Bonora:2002rn}
L.~Bonora, D.~Mamone and M.~Salizzoni,
{\it Vacuum string field theory ancestors of the GMS solitons},
JHEP {\bf 0301} (2003) 013
[arXiv:hep-th/0207044].

\bibitem{Bars:2001ag}
I.~Bars,
{\it Map of Witten's * to Moyal's *},
Phys.\ Lett.\ B {\bf 517} (2001) 436
[arXiv:hep-th/0106157].

\bibitem{Bars:2002nu}
I.~Bars and Y.~Matsuo,
{\it Computing in string field theory using the Moyal star product},
Phys.\ Rev.\ D {\bf 66} (2002) 066003
[arXiv:hep-th/0204260].

\bibitem{Douglas:2002jm}
M.~R.~Douglas, H.~Liu, G.~Moore and B.~Zwiebach,
{\it Open string star as a continuous Moyal product},
JHEP {\bf 0204} (2002) 022
[arXiv:hep-th/0202087].

\bibitem{Kluson:2001sb}
J.~Kluso\v{n},
{\it Some remarks about Berkovits' superstring field theory},
JHEP {\bf 0106} (2001) 045
[arXiv:hep-th/0105319].

\bibitem{Marino:2001ny}
M.~Mari\~{n}o and R.~Schiappa,
{\it Towards vacuum superstring field theory: The supersliver},
J.\ Math.\ Phys.\  {\bf 44} (2003) 156
[arXiv:hep-th/0112231].

\bibitem{Arefeva:2002mb}
I.~Ya.~Aref'eva, D.~M.~Belov and A.~A.~Giryavets,
{\it Construction of the vacuum string field theory on a non-BPS brane},
JHEP {\bf 0209} (2002) 050
[arXiv:hep-th/0201197].

\bibitem{Berkovits:1995ab}
N.~Berkovits,
{\it Super-Poincar\'{e} invariant superstring field theory},
Nucl.\ Phys.\ B {\bf 450} (1995) 90, Erratum ibid.\ B {\bf 459} (1996) 439,
[arXiv:hep-th/9503099].

\bibitem{Witten:1986qs}
E.~Witten,
{\it Interacting field theory of open superstrings},
Nucl.\ Phys.\ B {\bf 276} (1986) 291.

\bibitem{Arefeva:1989cp}
I.~Ya.~Aref'eva, P.~B.~Medvedev and A.~P.~Zubarev,
{\it New representation for string field solves the consistence problem for open superstring field},
Nucl.\ Phys.\ B {\bf 341} (1990) 464.

\bibitem{Berkovits:2000hf}
N.~Berkovits, A.~Sen and B.~Zwiebach,
{\it Tachyon condensation in superstring field theory},
Nucl.\ Phys.\ B {\bf 587} (2000) 147
[arXiv:hep-th/0002211].

\bibitem{DeSmet:2000dp}
P.~J.~De Smet and J.~Raeymaekers,
{\it Level four approximation to the tachyon potential in superstring field  theory},
JHEP {\bf 0005} (2000) 051
[arXiv:hep-th/0003220].

\bibitem{Aref'eva:2000mb}
I.~Ya.~Aref'eva, A.~S.~Koshelev, D.~M.~Belov and P.~B.~Medvedev,
{\it Tachyon condensation in cubic superstring field theory},
Nucl.\ Phys.\ B {\bf 638} (2002) 3
[arXiv:hep-th/0011117].

\bibitem{Kluson:2002kk}
J.~Kluso\v{n},
{\it Some solutions of Berkovits' superstring field theory},
arXiv:hep-th/0201054.

\bibitem{Arefeva:2002sg}
I.~Ya.~Aref'eva, A.~A.~Giryavets and A.~S.~Koshelev,
{\it NS ghost slivers},
Phys.\ Lett.\ B {\bf 536} (2002) 138
[arXiv:hep-th/0203227].

\bibitem{Ohmori:2002ah}
K.~Ohmori,
{\it Comments on solutions of vacuum superstring field theory},
JHEP {\bf 0204} (2002) 059
[arXiv:hep-th/0204138].

\bibitem{Ohmori:2002kj}
K.~Ohmori,
{\it On ghost structure of vacuum superstring field theory},
Nucl.\ Phys.\ B {\bf 648} (2003) 94
[arXiv:hep-th/0208009].

\bibitem{Ohmori:2003vq}
K.~Ohmori,
{\it Level-expansion analysis in NS superstring field theory revisited},
arXiv:hep-th/0305103.

\bibitem{Berkovits:1997pq}
N.~Berkovits and W.~Siegel,
{\it Covariant field theory for self-dual strings},
Nucl.\ Phys.\ B {\bf 505} (1997) 139
[arXiv:hep-th/9703154].

\bibitem{Gross:1986ia}
D.~J.~Gross and A.~Jevicki,
{\it Operator formulation of interacting string field theory},
Nucl.\ Phys.\ B {\bf 283} (1987) 1.

\bibitem{Ohta:wn}
N.~Ohta,
{\it Covariant Interacting String Field Theory In The Fock Space
Representation},
Phys.\ Rev.\ D {\bf 34} (1986) 3785,
Erratum ibid.\ D {\bf 35} (1987) 2627.

\bibitem{Belov:2002fp}
D.~M.~Belov,
{\it Diagonal representation of open string star and Moyal product},
arXiv:hep-th/0204164.

\bibitem{Gaiotto:2001ji}
D.~Gaiotto, L.~Rastelli, A.~Sen and B.~Zwiebach,
{\it Ghost structure and closed strings in vacuum string field theory},
Adv.\ Theor.\ Math.\ Phys.\  {\bf 6} (2003) 403
[arXiv:hep-th/0111129].

\bibitem{Berkovits:1994vy}
N.~Berkovits and C.~Vafa,
{\it N=4 topological strings},
Nucl.\ Phys.\ B {\bf 433} (1995) 123
[arXiv:hep-th/9407190].

\bibitem{Lechtenfeld:2002cu}
O.~Lechtenfeld, A.~D.~Popov and S.~Uhlmann,
{\it Exact solutions of Berkovits' string field theory},
Nucl.\ Phys.\ B {\bf 637} (2002) 119
[arXiv:hep-th/0204155].

\bibitem{Friedan:1985ge}
D.~Friedan, E.~J.~Martinec and S.~H.~Shenker,
{\it Conformal invariance, supersymmetry and string theory},
Nucl.\ Phys.\ B {\bf 271} (1986) 93.

\bibitem{Gross:1987pp}
D.~J.~Gross and A.~Jevicki,
{\it Operator formulation of interacting string field theory. 3. NSR Superstring},
Nucl.\ Phys.\ B {\bf 293} (1987) 29.

\bibitem{Gaberdiel:1997ia}
M.~R.~Gaberdiel and B.~Zwiebach,
{\it Tensor constructions of open string theories I: Foundations},
Nucl.\ Phys.\ B {\bf 505} (1997) 569
[arXiv:hep-th/9705038].

\bibitem{Kishimoto:2001ac}
I.~Kishimoto,
{\it Some properties of string field algebra},
JHEP {\bf 0112} (2001) 007
[arXiv:hep-th/0110124].

\bibitem{Bonora:2003xp}
L.~Bonora, C.~Maccaferri, D.~Mamone and M.~Salizzoni,
{\it Topics in string field theory},
arXiv:hep-th/0304270.

\bibitem{Kostelecky:2000hz}
V.~A.~Kosteleck\'{y} and R.~Potting,
{\it Analytical construction of a nonperturbative vacuum for the open bosonic string},
Phys.\ Rev.\ D {\bf 63} (2001) 046007
[arXiv:hep-th/0008252].

\bibitem{Mandelstam:jk}
S.~Mandelstam,
{\it Interacting string picture of dual resonance models},
Nucl.\ Phys.\ B {\bf 64} (1973) 205.

\bibitem{Mandelstam:1974fb}
S.~Mandelstam,
{\it Lorentz properties of the three - string vertex},
Nucl.\ Phys.\ B {\bf 83} (1974) 413.

\bibitem{Mandelstam:1974fq}
S.~Mandelstam,
{\it Dual - resonance models},
Phys.\ Rept.\  {\bf 13} (1974) 259.

\bibitem{Kaku:xu}
M.~Kaku and K.~Kikkawa,
{\it The field theory of relativistic strings. Pt.~2: Loops and pomerons},
Phys.\ Rev.\ D {\bf 10} (1974) 1823.

\bibitem{Cremmer:1974jq}
E.~Cremmer and L.~Gervais,
{\it Combining and splitting relativistic strings},
Nucl.\ Phys.\ B {\bf 76} (1974) 209.

\bibitem{Cremmer:1974ej}
E.~Cremmer and L.~Gervais,
{\it Infinite component field theory of interacting relativistic strings
and dual theory},
Nucl.\ Phys.\ B {\bf 90} (1975) 410.

\bibitem{LeClair:1988sp}
A.~LeClair, M.~E.~Peskin and C.~R.~Preitschopf,
{\it String field theory on the conformal plane. 1. Kinematical
principles},
Nucl.\ Phys.\ B {\bf 317} (1989) 411.

\bibitem{LeClair:1988sj}
A.~LeClair, M.~E.~Peskin and C.~R.~Preitschopf,
{\it String field theory on the conformal plane. 2. Generalized gluing},
Nucl.\ Phys.\ B {\bf 317} (1989) 464.

\bibitem{Okuyama:2002yr}
K.~Okuyama,
{\it Ghost kinetic operator of vacuum string field theory},
JHEP {\bf 0201} (2002) 027
[arXiv:hep-th/0201015].

\bibitem{Bordes:1993cd}
J.~Bordes, A.~Abdurrahman and F.~Anton,
{\it N string vertices in string field theory},
Phys.\ Rev.\ D {\bf 49} (1994) 2966
[arXiv:hep-th/9306029].

\bibitem{Ihl:200nie}
M.~Ihl, A.~Kling and S.~Uhlmann, to appear

\bibitem{Lechtenfeld:2000qj}
O.~Lechtenfeld and A.~D.~Popov,
{\it On the integrability of covariant field theory for open N = 2 strings},
Phys.\ Lett.\ B {\bf 494} (2000) 148
[arXiv:hep-th/0009144].

\bibitem{Kling:2002ht}
A.~Kling, O.~Lechtenfeld, A.~D.~Popov and S.~Uhlmann,
{\it On nonperturbative solutions of superstring field theory},
Phys.\ Lett.\ B {\bf 551} (2003) 193
[arXiv:hep-th/0209186].

\bibitem{Kling:2002vi}
A.~Kling, O.~Lechtenfeld, A.~D.~Popov and S.~Uhlmann,
{\it Solving string field equations: New uses for old tools},
Fortschr.\ Phys.\ {\bf 51} (2003) 775
[arXiv:hep-th/0212335].

\bibitem{Gross:1986fk}
D.~J.~Gross and A.~Jevicki,
{\it Operator formulation of interacting string field theory. 2},
Nucl.\ Phys.\ B {\bf 287} (1987) 225.

\bibitem{Samuel:1986wp}
S.~Samuel,
{\it The ghost vertex in E. Witten's string field theory},
Phys.\ Lett.\ B {\bf 181} (1986) 255.

\bibitem{Maccaferri:2003rz}
C.~Maccaferri and D.~Mamone,
{\it Star democracy in open string field theory},
JHEP {\bf 0309} (2003) 049
[arXiv:hep-th/0306252].

\end{thebibliography}
\end{document}